\documentclass[prb,aps,10pt,twocolumn,showpacs,floats,amsmath,amssymb,superscriptaddress]{revtex4-1}

\usepackage{hyperref}
\usepackage{graphicx}% Include figure files
\usepackage{dcolumn}% Align table columns on decimal point
\usepackage{bm}% bold math
\usepackage{amssymb}% for unusual symbols
\usepackage[latin1]{inputenc}
\usepackage{subfigure}
\usepackage{color}
\usepackage{xcolor}
\usepackage{enumerate}
\usepackage{ulem}   % pour barrer du texte

\def \be{\begin{equation}}
\def \ee{\end{equation}}

\definecolor{orange}{rgb}{1,0.5,0}

\begin{document}

\title{Nanowire-based thermoelectric ratchet in the hopping regime}

\author{Riccardo Bosisio}
%\footnote{Present address: NEST, Instituto Nanoscienze-CNR and Scuola Normale Superiore, I-56127 Pisa, Italy}}
%\email{riccardo.bosisio@nano.cnr.it}
\affiliation{SPIN-CNR, Via Dodecaneso 33, 16146 Genova, Italy}
\affiliation{NEST, Instituto Nanoscienze-CNR and Scuola Normale Superiore, I-56127 Pisa, Italy}

\author{Genevi\`eve Fleury}
\email{genevieve.fleury@cea.fr}
\affiliation{SPEC, CEA, CNRS, Universit\'e Paris-Saclay,\\
CEA Saclay, 91191 Gif-sur-Yvette Cedex, France}

\author{Jean-Louis Pichard}
%\email{jean-louis.pichard@cea.fr}
\affiliation{SPEC, CEA, CNRS, Universit\'e Paris Saclay,\\
CEA Saclay, 91191 Gif-sur-Yvette, France}

\author{Cosimo Gorini}
%\email{cosimo.gorini@physik.uni-regensburg.de}
%\footnote{Present address: Institut für Theoretische Physik, Universität Regensburg, 93040 Regensburg, Germany}}
\affiliation{Institut für Theoretische Physik, Universität Regensburg, D-93040 Regensburg, Germany}

\begin{abstract}
We study a thermoelectric ratchet consisting of an array of disordered nanowires  arranged in parallel on top of an insulating substrate, and contacted asymmetrically to two electrodes.
Transport is investigated in the Mott hopping regime, when localized electrons can propagate through the nanowires via thermally assisted hops. When the electronic temperature in the nanowires is different from the phononic one in the substrate, we show that a finite electrical current is generated even in the absence of driving forces between the electrodes.
We discuss the device performance both as an energy harvester, when an excess heat from the substrate is converted into useful power, and as a refrigerator, when an external power is supplied to cool down the substrate. 
\end{abstract}


\maketitle
\section{Introduction}

The first golden age of thermoelectricity dates back to Ioffe's suggestion in the 1950s of using semiconductors in thermoelectric 
modules.~\cite{Ioffe1957} In spite of sustained efforts, 
it only led to thermoelectric devices limited by their poor efficiency to niche applications. 
Interest in thermoelectricity was revived in the 1990s by nanostructuration and the appealing perspectives of enhanced efficiency 
it offers.~\cite{Hicks1993,Hicks1993bis}
Nowadays the idea of exploiting multi-terminal thermolectric setups is driving the field through
a new season of very intense activity~\cite{Saito2011,SanchezDavid2011,Horvat2012,Balachandran2013,
Brandner2013,Mazza2014,Bosisio2015,Whitney2013,Machon2013,Mazza2015,Valentini2015,Sanchez2011,Jordan2013,
Sothmann2013,Roche2015,Hartmann2015,Thierschmann2015,Hofer2015,Sanchez2015,Rutten2009,Ruokola2012,
Bergenfeldt2014,Cleuren2012,Mari2012,Entin2010,Jiang2013bis,Entin2015,Jiang2012,Jiang2015,Bosisio20142,BosisioGorini2015,Bosisio2015bis}.
In contrast with conventional two-terminal thermoelectrics, multi-terminal thermoelectrics aims at studying a conductor connected, 
in addition to the two reservoirs at its ends, to (at least) one other reservoir, 
be it a mere probe~\cite{Saito2011,SanchezDavid2011,Horvat2012,Balachandran2013,Brandner2013,Hofer2015}, 
a normal electronic reservoir~\cite{Whitney2013,Mazza2014,Bosisio2015}, a superconducting lead~\cite{Whitney2013,Machon2013,Mazza2015,Valentini2015}, or a reservoir of fermionic~\cite{Sanchez2011,Jordan2013,Sothmann2013,Sanchez2015,Hofer2015} or bosonic~\cite{Rutten2009,Ruokola2012,Bergenfeldt2014,Pekola2007,Cleuren2012,Mari2012,Entin2010,Jiang2013bis,Entin2015,Jiang2012,Jiang2015,Bosisio20142,BosisioGorini2015} nature that can only exchange energy with the system. Investigations carried out so far have shown that the multi-terminal geometry has generally a positive impact on the performance of the thermoelectric devices~\cite{Balachandran2013,Hofer2015,Mazza2014,Entin2015,Mazza2015}, compared to their two-terminal counterparts. It also opens up new perspectives, such as the possibility of implementing a magnetic thermal switch~\cite{Bosisio2015} or of separating and controlling heat and charge flows independently~\cite{Mazza2015}.\\
\indent In the following we focus on three-terminal thermoelectric harvesters, which can be also viewed as three-terminal thermoelectric ratchets using excess heat coming from the environment to generate a directed electrical current through the conductor.
The dual cooling effect, enabling to cool down the third terminal by investing work from voltage applied across the conductor, is also studied.
One of the first proposed realizations of three-therminal thermoelectric harvester was a Coulomb-blockaded quantum dot~\cite{Sanchez2011} 
exchanging thermal energy with a third \textit{electronic} bath, capacitively coupled. 
Its feasibility has been recently confirmed experimentally,~\cite{Roche2015, Hartmann2015} 
though the output power turns out to be too small for practical purposes.  
Since then, other quantum dot- or quantum well-based devices have been put forward \cite{Jordan2013,Sothmann2013}
with the hope of overcoming the problem. On the other hand, various devices running on energy exchanges with a third
\textit{bosonic} reservoir have been discussed.
%Besides magnons~\cite{Sothmann2012bis} and photons,\cite{Ruokola2012,Pekola2007,Rutten2009,Cleuren2012,Mari2012,Bergenfeldt2014} phonons~\cite{Entin2010,Jiang2013bis,Jiang2012,Entin2014,Jiang2015} are possible vehicles to envisage for powering thermoelectric energy harvesters or refrigerators. 
In particular, phonon-driven mechanisms have been considered at a theoretical level in two-levels systems or chains of localized states along nanowires (NWs) in the context of phonon-assisted hopping transport.\cite{Jiang2012,Jiang2013,Jiang2015}
More generally, NW-based devices have been at the heart of experimental studies
on future thermoelectrics for over a decade \cite{Li2012,Kim2013}.  Two critical advantages of such setups are nanostructuration
\cite{Hicks1993bis,Nakpathomkun2010,Hochbaum2008} and scalability,~\cite{Persson2009,Wang2009,Curtin2012,Farrell2012,Stranz2013}
the latter being a crucial requirement for substantial output power.  Furthermore, NWs are core products
of the semiconductor industry, commonly fabricated up to large scales and used in a broad range of applications,
from thermoelectrics to photovoltaics~\cite{Garnett2011,Lapierre2013} or biosensing.\cite{Chen2011}\\
\indent In light of the above, the energy harvester/cooler we propose is a NW-based 
three-terminal thermoelectric ratchet, as sketched in Fig.~\ref{fig:sys}.  
A set of disordered (doped) semiconductor NWs is connected in parallel
to two electronic reservoirs and deposited on a substrate.  The electronic states in the NWs are localized by disorder,
but transport is possible thanks to phonons from the substrate, which allow activated hops between the localized states.\cite{Mott1969,Ambegaokar1971} 
In two recent works~\cite{Bosisio20142,BosisioGorini2015} we showed that similar setups exhibit remarkable \textit{local} (two-terminal)
thermoelectric properties. This is mainly because in the hopping regime the transport energy window around the Fermi level
is much larger than the thermal energy, \textit{i.e.} the one of conventional band transport,
making it possible to exploit particle-hole asymmetry across a wide energy range. Therefore the thermopower which is a direct measure of the "degree" of particle/hole asymmetry can possibly reach large values, somehow compensating -- regarding thermoelectric performance -- the smallness of the electrical conductance in the hopping regime.\\
\indent In the present manuscript, we explore the potential of our device in the phonon-assisted activated regime for \textit{non-local} thermoelectric conversion (the core idea behind multi-terminal thermoelectrics). More precisely,
we are mainly interested in harvesting waste heat from a hot substrate (the third terminal) to generate an electric current
between the two electronic reservoirs, thus supplying a load.  
The process requires to define ratchet pawls forcing charge carriers to escape the NWs 
preferably on one side.  Quite generally, this can be achieved by breaking spatial mirror symmetry (see \textit{e.g.} Refs.~\onlinecite{Rahman2006,Linke1999,Sassine2008}).
Particle-hole symmetry need be broken as well, which is the basic requirement for any thermoelectric device.
Both symmetry-breaking conditions are implemented by inserting different energy filters at the left and right metal-semiconductor contacts.
Two simple models of energy filter mimicking a Schottky barrier and an open quantum dot are discussed in detail.  
The thermoelectric ratchet power factor $Q$, characterizing its output power in the heat engine configuration,
and the electronic figure of merit $ZT$, controlling its efficiency in the absence of parasitic phonon contribution,
depend on the choice of contact type and the degree of asymmetry.  Remarkably, both quantities reach maximum values 
in the same range of parameters, i.e. large values $ZT\gg 1$ at high (scalable) output powers $Q$ can be obtained.  
In all respects, the three-terminal non-local thermoelectric converter 
is found to be much more performant that the corresponding local, two-terminal one.
Besides waste heat harvesting, we also briefly study the refrigerator configuration, in which a current flowing in the NWs 
can be used to cool down the phonon bath.\\
\indent The outline is as follows. In Sec.~\ref{sec:model}, we describe the model and the (numerical) method used to calculate the currents and the thermoelectric coefficients. In Sec.~\ref{sec:conversion}, we discuss different implementations of ratchet pawls at the metal-semiconductor contacts and show the ratchet effect \textit{i.e.} the conversion of excess heat from the substrate into a directed electrical current. Sec.~\ref{sec:results} is dedicated to the estimation of the device performance.  We conclude in Sec.~\ref{ccl}. Two appendices are added to discuss additional results.

\section{Model and Method}
\label{sec:model}
\begin{figure}[b]
    \includegraphics[clip,keepaspectratio,width=\columnwidth]{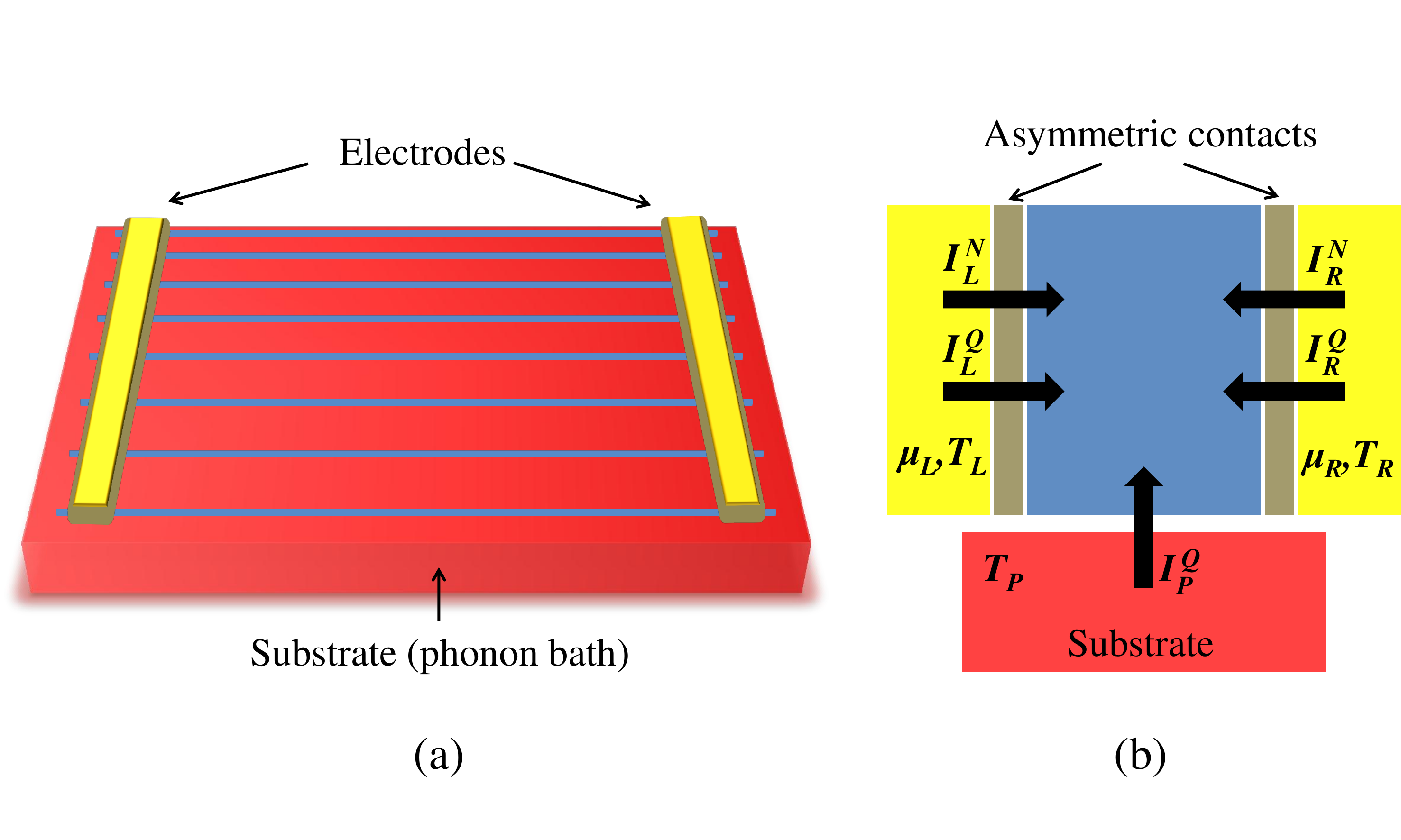}
    \caption{(Color online) (a) Sketch of the NW-based ratchet studied in this work. A large array of parallel disordered NWs (in blue) is deposited onto an insulating substrate (in red) acting as a phonon bath. The NWs are attached to metallic electrodes (yellow) via asymmetric contacts. (b) Schematic of the three-terminal transport. Electrons inside the NWs are localized within their impurity band (blue) and can exchange phonons with the substrate (held at temperature $T_P$). Under condition of asymmetric couplings (grey strips) to the electrodes (yellow), this can generate net currents flowing through the NWs. The particle and heat currents between the NWs and the three reservoirs are represented by the arrows, assuming they are positive when they enter the system.}
   \label{fig:sys}
\end{figure}

Phonon-activated transport through the NW-based ratchet [Fig.~\ref{fig:sys}(a)] is described in linear response,
and thus characterized by a three-terminal Onsager matrix. The latter is defined in Sec.~\ref{subsecOnsager}.
The way it is computed, by solving the random resistor network problem, is briefly reviewed in Sec.~\ref{subsechophop}.

\subsection{Onsager formalism for the three-terminal thermoelectric device}
\label{subsecOnsager}
We consider a conducting region connected to two electronic reservoirs $L$ and $R$ at equilibrium,
characterized by electrochemical potentials $\mu_L$, $\mu_R$, and temperatures $T_L$, $T_R$, and to a bosonic reservoir $P$ 
at temperature $T_P$ [see Fig.~\ref{fig:sys}(b)].
Heat and particles can be exchanged with $L$ and $R$, but only heat with $P$. 
The particle currents $I^N_L$, $I^N_R$, and the heat currents $I^Q_L$, $I^Q_R$, $I^Q_P$ are defined positive when entering
the conducting region from the reservoir $\alpha=L,R,P$. 
The right terminal $R$ is chosen as reference ($\mu_R\equiv\mu$ and $T_R\equiv T$) and we set
\be
\delta\mu = \mu_L-\mu_R,~~\delta T=T_L-T_R,~~\mathrm{and}~~\delta T_P=T_P-T_R\,.
\ee
In linear response the independent currents $I^N_L$, $I^Q_L$, and $I^Q_P$ are expressed \`a la Onsager
 in terms of the corresponding driving forces
\be
\begin{pmatrix}
I^N_L \\
I^Q_L \\
I^Q_P
\end{pmatrix}
=
\begin{pmatrix}
L_{11} & L_{12} & L_{13} \\
L_{12} & L_{22} & L_{23} \\
L_{13} & L_{23} & L_{33}
\end{pmatrix}
\begin{pmatrix}
\delta\mu / T \\
\delta T / T^2 \\
\delta T_P / T^2
\end{pmatrix}\,.
\label{eq_onsager}
\ee
In writing the above we have exploited the Casimir-Onsager relations\cite{Callen1985} $L_{ij}=L_{ji}$ for $i\neq j$,
valid in the absence of time-reversal symmetry breaking.

In the following we focus on the specific case $\delta T=0$, that is, the system and the electronic reservoirs share the same
temperature $T$. On the other hand, we consider $\delta\mu, \delta T_P\neq0$ and discuss several possibilities offered by this setup 
in terms of energy harvesting (when the heat provided by the phonon bath is exploited to produce electrical work) 
and cooling (when an external work is invested to cool down the phonon bath).

\subsection{Nanowire array in the phonon-assisted activated regime}
\label{subsechophop}
The device [see Fig.~\ref{fig:sys}(a)] is realized by depositing a set of $M$ disordered NWs in parallel onto an insulating
substrate (which plays the role of a phonon bath), and connecting them asymmetrically to two metallic electrodes
(acting as electron reservoirs). The electrodes are assumed to be thermally isolated from the substrate (not highlighted in Fig.~\ref{fig:sys}) such that the electron and phonon reservoirs can be held at different temperatures. Each NW is modeled as a one-dimensional wire of length 
$L=Na$, with $a$ the average nearest-neighbor distance (set equal to one from here on).
Disorder localizes the $N$ electronic states, assumed uniformly distributed in space and energy within an impurity band $[-2\epsilon,2\epsilon]$ ($\epsilon$ is the energy unit)
with constant density of states $\nu=1/(4\epsilon)$ and constant localization length $\xi$.
In each NW no site can be doubly occupied due to Coulomb repulsion, but we otherwise neglect interactions.  
Also, we assume the NWs to be independent, i.e. no \textit{inter}-wire hopping is considered.
This setup was extensively discussed in 
our previous works~\cite{Bosisio20142, BosisioGorini2015,Bosisio2015bis}, to which we refer for more details. 
Note that in Refs.~\onlinecite{Bosisio20142, BosisioGorini2015,Bosisio2015bis} the contacts were symmetric, and both $\nu$ and $\xi$
were chosen energy-dependent to infer band-edge properties.\footnote{In previous works~\cite{Bosisio20142,BosisioGorini2015,Bosisio2015bis}
focusing on band-edge transport, the energy dependence of the localization length was crucial and thus taken into account.  
Such dependence is here largely inconsenquential: apart from a brief discussion of the less relevant configuration of 
Fig.~\ref{fig:ratchet}(a1), (a2), (a3), the band edges will not be probed.}\\
\indent Electrons tunnel between reservoir $\alpha=R,L$ and the $i$-th localized state in a given NW at the rate
(Fermi Golden Rule)
\be
\Gamma_{i\alpha}=\gamma_{i\alpha}(E_i)f_i[1-f_\alpha(E_i)],
\label{eq_rates1}
\ee
where $f_i$ is the occupation probability of state $i$ and $f_\alpha(E)=[\exp((E-\mu_\alpha)/k_BT_\alpha)+1]^{-1}$
the Fermi distribution in reservoir $\alpha$.  State $i$ is coupled to reservoir $\alpha$ via
$\gamma_{i\alpha}(E_i)=\gamma_{e\alpha}(E_i)\exp(-2x_{i\alpha}/\xi)$,
with $\gamma_{e\alpha}$ and $x_{i\alpha}$ respectively its coupling with and distance to 
the latter.
Propagation through the NW takes place via (inelastic) phonon-assisted hops.\cite{Bosisio20142,BosisioGorini2015,Jiang2012,Jiang2013}
The transition rate between states $i$ and $j$, at energies $E_i$ and $E_j$, is
\be
\Gamma_{ij}=\gamma_{ij} f_i (1-f_j) [N_{ij}+\Theta(E_j-E_i)],\label{eq_rates2}
\ee
where $N_{ij}=[\exp(|E_j-E_i|/k_BT_P)-1]^{-1}$ is the probability of having a phonon with energy $|E_j-E_i|$, 
$\Theta$ is the Heaviside function, $\gamma_{ij}=\gamma_{ep}\exp(-2x_{ij}/\xi)$, $x_{ij}=|x_i-x_j|$ 
and $\gamma_{ep}$ is the electron-phonon coupling.\\
\indent The particle and heat currents through the $k$-th NW are computed in linear response by solving the random resistor network problem\cite{Miller1960,Ambegaokar1971} (for recent reviews within the framework of thermoelectric transport, see \textit{e.g.} Refs.~\onlinecite{Bosisio20142,Jiang2013}). The method yields the non-equilibrium steady-state occupation
probabilities $f_i$, and thus the transition rates \eqref{eq_rates1} and~\eqref{eq_rates2}.  The particle currents
between state $i$ in NW $k$ and reservoir $\alpha$, and between each pair of localized states $i,j$, read
\begin{align}
&I^{(k)}_{ij}= \Gamma^{(k)}_{ij}-\Gamma^{(k)}_{ji},\cr
&I^{(k)}_{i\alpha}=\Gamma^{(k)}_{i\alpha}-\Gamma^{(k)}_{\alpha i},
\end{align}
%
%where the superscript $k$ indicates that we are referring to the current contributions in the $k$-th nanowire.
whereas the total particle and heat currents through NW $k$ are
\begin{align}
&I^{N(k)}_\alpha = \sum_i I^{(k)}_{\alpha i},\cr
&I^{Q(k)}_\alpha = \sum_i I^{(k)}_{\alpha i}(E^{(k)}_i-\mu_\alpha),\cr
&I^{Q(k)}_P = \frac{1}{2}\sum_{i\neq j} I^{(k)}_{ij}(E^{(k)}_j-E^{(k)}_i).
\end{align}
The total currents flowing through the whole device are given by summing over all $M$ NWs in the array:
%
%\begin{align}
\be
\label{eq_IN}
I^N_\alpha = \sum_k I^{N(k)}_\alpha,~~ I^Q_\alpha = \sum_k I^{Q(k)}_\alpha,~~ I^Q_P = \sum_{k} I^{Q(k)}_P.
\ee
%\end{align}
%
Charge and energy conservation respectively implies
\be
\label{eq_cons2}
I^N_L=-I^N_R\,,
\ee
\be
\label{eq_cons}
I^Q_L+I^Q_R+I^Q_P=-\,\delta\mu\, I^N_L\,,
\ee
where $-\,\delta\mu\, I^N_L$ is the dissipated (Joule) heat.
Notice that, by virtue of Eq.\,\eqref{eq_onsager}, calculating the currents by imposing only one driving force and setting the other to zero allows us to compute
one column of the Onsager matrix. Upon iterating this procedure for $\delta\mu$, $\delta T$ and $\delta T_P$, the full matrix can be built up.

\subsection{Parameters setting}

To get rid of the disorder-induced fluctuations of $I^{N(k)}_\alpha$, $I^{Q(k)}_\alpha$, and $I^{Q(k)}_P$, we take a sufficiently large number 
$M=10^4$ of parallel NWs [and up to $M=2.10^{5}$ for data in Fig.\ref{fig:ratchet}(A)]. Thereby, all quantities plotted in figures hereafter self-average.
Moreover, throughout the paper, we set the NW length to $N=100$ and the localization length to $\xi=4$. The temperature is also fixed to $k_BT=0.5\,\epsilon$,
so as to be in the activated regime\footnote{It is indeed close to the Mott temperature $k_BT_M=2/\xi\nu=2\,\epsilon$ and much larger than the activation temperature $k_BT_x=\xi/(2\nu N^2)=8.10^{-4}\,\epsilon$ (see Ref.\onlinecite{Bosisio20142} for more details).}. For completeness, temperature effects are discussed in Appendix~\ref{app_T}.  
For this set of parameters, the Mott hopping energy \textit{i.e.} the range of energy states effectively contributing to transport,  is\cite{Bosisio20142} $\Delta\simeq\sqrt{2k_BT/(\xi\nu)}=\epsilon$.
Having fixed a specific size is no limitation: as we discussed in a recent work\cite{BosisioGorini2015} the transport coefficients are basically independent of the NW size in the activated regime. Also, since $\gamma_{ep}$ is weakly dependent on the $E_i$'s and $x_{ij}$'s compared to the exponential factors in Eq.\eqref{eq_rates2}, we take it constant. And since the variables $f_i$ are only functions of the couple $(\gamma_{eL}/\gamma_{ep}, \gamma_{eR}/\gamma_{ep})$ and not of the three parameters $\gamma_{eL}$, $\gamma_{eR}$, and $\gamma_{ep}$, we choose $\gamma_{ep}=\epsilon/\hbar$ without loss of generality.

\section{Heat to charge conversion}
\label{sec:conversion}

In this section we discuss how to exploit the temperature difference $\delta T_P$ between NW electrons and
substrate phonons to generate a net particle (charge) current in the absence of any voltage bias ($\delta\mu=0$). 
To convert heat coming from the substrate to charge current, two requirements are needed: 
\begin{enumerate}[(i)]
\item broken left $\leftrightarrow$ right inversion symmetry (here due to different left and right contacts);
\item broken electron-hole symmetry.
\end{enumerate}
In the original Feynman's brownian ratchet\cite{Feynman1964}, condition (i) is guaranteed by the presence of pawls attached to the paddle wheel preventing one rotation direction of the wheel. In our work, condition (i) ensures that electron- and hole-excitations -- created around $\mu$ within each NW when the system is driven out of equilibrium by $\delta T_P$ -- preferably escape on one side. Condition (ii) is required as well since here the investigated ratchet effect relies on a thermoelectric effect. Indeed, if (i) is satisfied but not (ii), the contribution of the electron-like particles above $\mu$ and the contribution of the hole-like particles below $\mu$ compensate each other on average, though each contribution taken separetely is non-zero by virtue of (i).
Hereafter, we discuss different implementations of conditions (i) and (ii), and show evidence of the thermoelectric three-terminal ratchet effect in our setup. The effect is illustrated in Fig.~\ref{fig:ratchet} for different kinds of asymmetric contacts.

\subsection{Asymmetric contacts as ratchet pawls}
\label{sec3A}

\begin{figure*}
    \includegraphics[clip,keepaspectratio,width=\textwidth]{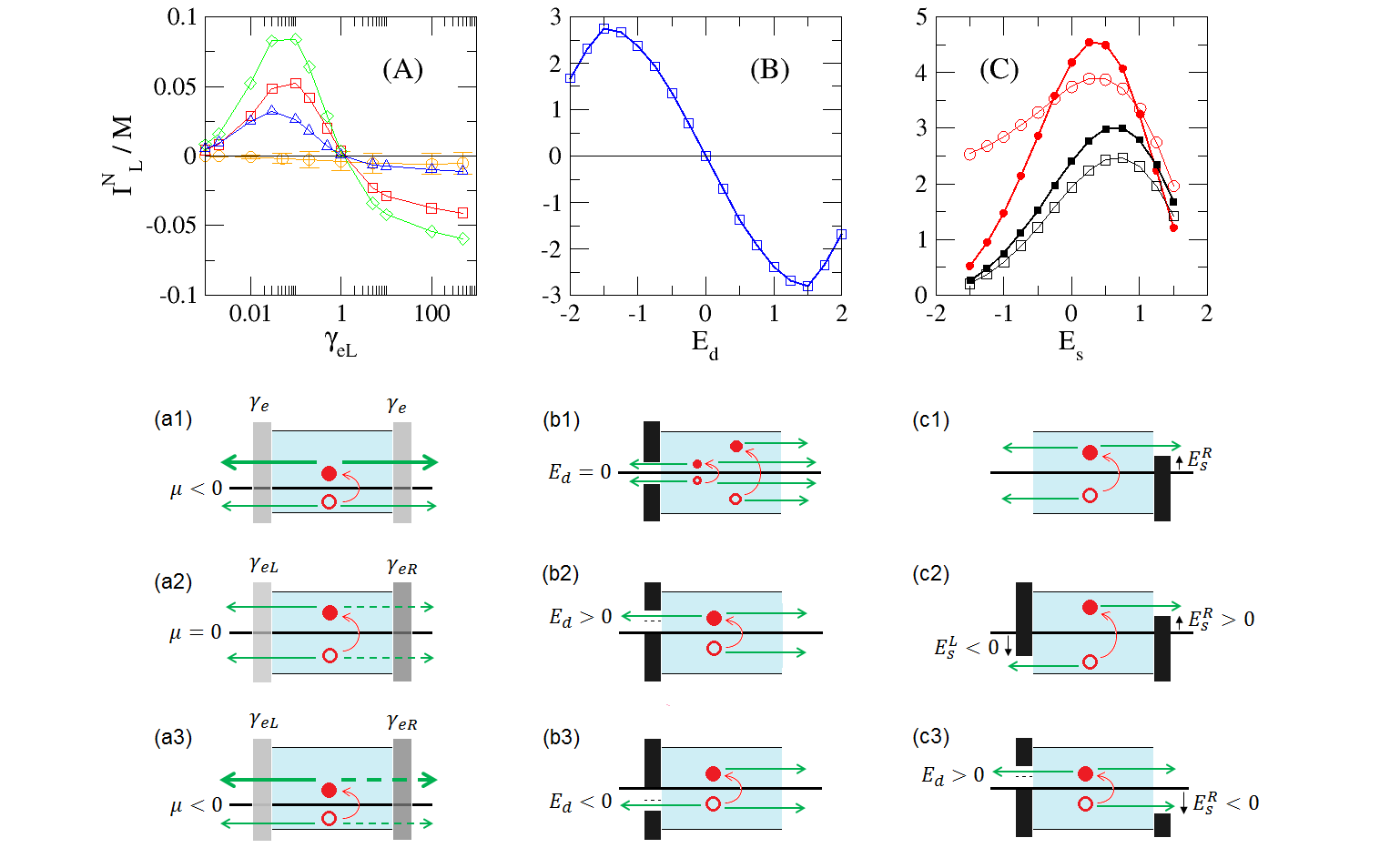}
    \caption{(Color online) Illustration of the ratchet effect powered by $\delta T_P> 0$, for various types of contacts. The total particle current $I^N_L$ (see Eq.\eqref{eq_IN}), obtained with $\delta T_p=10^{-3}\,\epsilon$ and $\delta\mu=0$, is plotted (in units of $10^5 \epsilon/\hbar$ and divided by $M$) in top panels (A-C) for different configurations shown below (a1-c3) and listed at the end of Sec.\ref{sec3A}. (Left) Energy-independent tunnel barriers. If the contacts are symmetric (a1) or if $\mu=0$ (a2), $I^N_L=0$ on average. If both symmetries are broken (a3) a net current is generated. (A) $I^N_L/M$ as a function of $\gamma_{eL}$ (in units of $\epsilon/\hbar$) for fixed $\gamma_{eR}=\epsilon/\hbar$ and various positions of $\mu$ ($\mu/\epsilon=0$ ({\large{\color{orange}$\circ$}}), $-0.5$ ({\scriptsize{\color{red}$\square$}}), $-1$ ({\large{\color{green}$\diamond$}}) and $-2.5$ ({\scriptsize{\color{blue}$\triangle$}})). (Middle) Energy filter. When $E_d=0$ electron-hole symmetry makes the total current vanish (b1), whereas $E_d>0$ (b2) and $E_d<0$ (b3) correspond to negative and positive current, respectively. (B) $I^N_L/M$ at $\mu=0$, for constant $\gamma_{eR}=\epsilon/\hbar$ on the right and  an energy filter on the left with $\gamma_e=\epsilon/\hbar$, opening $\Gamma=\epsilon$, and tunable $E_d$ (in units of $\epsilon$). (Right) (c1) Single barrier configuration. The Schottky barrier prevents electron- and hole-like excitations with energy below $E_s^R$ to escape to the right contact. (c2) Double barrier configuration. A low energy filter is added at the left contact, forbidding electron- and hole-like excitations with energy above $E_s^L$ to tunnel leftward. (c3) Hybrid configuration. A Schottky barrier on the right is combined with an energy filter on the left. (C) $I^N_L/M$ at $\mu=0$, as a function of the barrier height $E_s$ (in units of $\epsilon$), with $\gamma_{e}=\epsilon/\hbar$, for the single barrier ({\tiny{$\blacksquare$}}, $E_s\equiv E_s^R$), double barrier ({\scriptsize{\color{red}$\bullet$}}, $E_s\equiv E_s^R = -E_s^L$), and hybrid cases ($E_s\equiv E_s^R$) with $E_d=0$ ({\scriptsize{$\square$}}) and $E_d=-\epsilon$ ({\large{\color{red}$\circ$}}).}
   \label{fig:ratchet}
\end{figure*}

Condition (i) is implemented by inserting different contacts at the left and right NW extremities.
Within the theoretical framework reviewed above, the contact between the NWs and the reservoir $\alpha$ is characterized by the coupling $\gamma_{e\alpha}$. We focus on some specific choices for this metal-semiconductor contact:
\begin{itemize}
\item $\gamma_{e\alpha}(E)=\gamma_{e\alpha}$ independent of the energy. This model, implemented via energy-independent tunnel barriers between NWs and electrodes, does not break electron-hole symmetry [condition (ii)].
However, this can be easily done by putting the device in a field effect transistor configuration\cite{Bosisio20142,BosisioGorini2015,Bosisio2015bis}, so as to have $\mu\neq 0$ (see Fig.~\ref{fig:ratchet}, left column).
\item $\gamma_{e\alpha}(E)=\gamma_{e\alpha}\Theta(E-E_s^\alpha)$, with $\Theta$ the Heaviside (step) function.  
This is the simplest model for a Schottky barrier at the NW-reservoir $\alpha$ interface,  
acting as an \textit{high energy filter} -- only charge carriers with energy $E_i$ above a certain threshold $E_s^\alpha$ can flow 
[see Fig.~\ref{fig:ratchet}~(c1)]. The barrier guarantees that condition (ii) is fulfilled.
\item $\gamma_{e\alpha}(E)=\gamma_{e\alpha}[1-\Theta(E-E_s^\alpha)]$, i.e. a simple model for a \textit{low energy filter} 
-- only charge carriers with energy $E_i$ below a certain threshold $E_s^\alpha$ can flow [see Fig.~\ref{fig:ratchet}~(c2)].  
Despite being more difficult to implement in practice, it offers an instructive toy model. Just as in the previous case, the barrier ensures that condition (ii) is satisfied.
\item $\gamma_{e\alpha}(E)=\gamma_{e\alpha}$ if $E\in[E_d-\Gamma/2,E_d+\Gamma/2]$ and $0$ elsewhere. 
This is a simple model for an \textit{energy filter}, allowing only charge carriers with energies $E$ inside a window $\Gamma$ around $E_d$
to flow into/out of the NWs (see Fig.~\ref{fig:ratchet}, middle column).
In practice, it could be realized by embedding a single level quantum dot in each NW close to electrode $\alpha$;  
$\Gamma$ would represent the dot opening, and $E_d$ its energy level, easy tunable with an external gate\cite{Nakpathomkun2010}. Even for $\mu=0$ at the band center, this model fulfills condition (ii) if $E_d\neq 0$. Moreover, for a large opening $\Gamma$ of the dot and a proper tuning of $E_d$, this model can mimic a low energy filter.
%\modif{May be seen as the quantum dot model with large opening $\Gamma$.}
\end{itemize}
To fulfill requirement (i), it is necessary to introduce different coupling functions $\gamma_{eL}(E)\neq \gamma_{eR}(E)$ to the left and right reservoirs. Hereafter, we consider the following asymmetric configurations:
\begin{enumerate}
\item ``Asymmetric tunnel contacts'': this is implemented by fabricating different energy-independent contacts $\gamma_{eL}\neq\gamma_{eR}$ [Fig.~\ref{fig:ratchet}(a3)].
\item ``Single filter'': an energy filter -- with $\gamma_{e\alpha}(E)=\gamma_{e}$ if $E\in[E_d-\Gamma/2,E_d+\Gamma/2]$ and $0$ elsewhere -- is placed on the left, and an energy-independent tunnel barrier $\gamma_{eR}(E)=\gamma_{e}$ on the right [Fig.~\ref{fig:ratchet}(b2-b3)].
\item ``Single barrier'': we consider an energy-independent tunnel barrier on the left $\gamma_{eL}(E)=\gamma_{e}$ and a Schottky barrier between the NWs and the right contact, $\gamma_{eR}(E)=\gamma_{e}\Theta(E-E_s^{R})$ [Fig.~\ref{fig:ratchet}(c1)].
\item ``Double barrier'': we consider a low energy filter on the left, $\gamma_{eL}(E)=\gamma_{e}[1-\Theta(E-E_s^{L})]$,  and a Schottky barrier on the right, $\gamma_{eR}(E)=\gamma_{e}\Theta(E-E_s^{R})$ [Fig.~\ref{fig:ratchet}(c2)].
\item ``Hybrid configuration'': as the previous one, but with the left low energy filter replaced by an energy filter (an embedded quantum dot), with $\gamma_{eL}(E)=\gamma_{e}$ if $E\in[E_d-\Gamma/2,E_d+\Gamma/2]$ and $0$ elsewhere [Fig.~\ref{fig:ratchet}(c3)]. This model is introduced as a refinement of the double barrier one, easier to implement experimentally.
\end{enumerate}

\subsection{Ratchet-induced charge current}

Once the phonon bath is heated up ($\delta T_P> 0$), the NW electrons are driven out of equilibrium and electron-hole excitations are created around $\mu$ (see the corresponding sketches in Fig.~\ref{fig:ratchet}).
Knowing that the couplings $\gamma_{i\alpha}(E_i)=\gamma_{e\alpha}(E_i)\exp(-2x_{i\alpha}/\xi)$ and that the electronic states are uniformly distributed along the NWs, we focus for simplicity on a single excitation at the NW center and discuss the phenomenology leading to a finite charge current generation in the different situations shown in Fig.~\ref{fig:ratchet} (the same reasoning can be extended on statistical grounds to the set of $N$ states inside each NW). Our qualitative predictions (based on the simplified pictures sketched in the bottom panels of Fig.~\ref{fig:ratchet}) are confirmed by the numerical simulations which take all excitations into account (top panels of Fig.~\ref{fig:ratchet}).\\
\indent The first column refers to the case of energy independent coupling factors $\gamma_{eL}$ and $\gamma_{eR}$. 
If they are equal ($\gamma_{eL}=\gamma_{eR}$) both electron- and hole-like excitations have the same probability 
to tunnel out to the left/right reservoir, and no net current flows [Fig.~\ref{fig:ratchet}(a1)]. 
Breaking this symmetry induces a preferential direction for tunneling out of the NW, 
thus fulfilling condition (i) [Fig.~\ref{fig:ratchet}(a2)].
However, without electron-hole symmetry breaking [condition (ii)], the number of hole-like excitations with energy $\mu-E$
equals on average\footnote{It is worth to stress that if we consider a single NW, electron-hole symmetry may be broken even at $\mu=0$ due to disorder; however, when considering a large set of NWs having constant density of states $\nu$, symmetry is restored on average.}
that of electron-like ones with energy $\mu+E$, resulting again in a vanishing current.
This second symmetry can be broken by shifting the electrochemical potential $\mu$ within the NWs impurity band via a top/back gate\cite{Bosisio20141,Bosisio20142} leading to $\mu\neq 0$. In this case, and provided $\gamma_{eL}\neq\gamma_{eR}$, 
a net current flows through the NW array [Fig.~\ref{fig:ratchet}(a3)].
The total particle current per NW $I^N_L/M$ is shown in panel (A) as function of the left coupling $\gamma_{eL}$ for fixed $\gamma_{eR}$ and different positions of $\mu$ in the impurity band.  Our qualitative analysis is confirmed: the current vanishes for $\mu=0$ at the band center
(at least within the error bars\footnote{They give an estimation of the difference between data obtained for finite $M$ and the quantity $I^N_L/M$ for $M\to\infty$.}), whereas it is non-zero for $\mu\neq 0$ once $\gamma_{eL}\neq\gamma_{eR}$. The sign of $I^N_L$ (taken positive when the flow of electrons goes from left to right) is given by the sign of $\mu(\gamma_{eL}/\gamma_{eR}-1)$.\\
\indent The middle column describes the effect of an energy filter of width $\Gamma$ and coupling $\gamma_e$, centered at $E_d$, and placed between the NWs and the left reservoir.
In this case $\mu=0$ is fixed at the impurity band center, and at the right contact $\gamma_{eR}(E)=\gamma_{e}$.
From the sketch we see that condition (i) is straightforwardly satisfied, whereas condition (ii) is fulfilled only if $E_d\neq 0$.
Interestingly, it is possible to control the direction of the current by simply adjusting the position of $E_d$: if $E_d>0$ [Fig.~\ref{fig:ratchet}(b2)] the electron-like excitation created above $\mu$ (within $\Gamma$) can tunnel left or right with equal probability, whereas the hole-like one (within $\Gamma$) can only escape to the right.
Other electron- and hole-excitations beyond the energy range $\Gamma$ around $\mu$ are equally coupled to the electrodes and do not contribute to the current.
A net electric current is thus expected to flow leftward ($I^N_L<0$).
By the same token, when $E_d<0$ [Fig.~\ref{fig:ratchet}(b3)] the hole-like excitation (within $\Gamma$) does not contribute, whereas the electron-like one can tunnel right: we thus expect a finite current flowing rightward ($I^N_L>0$). All these predictions are confirmed in panel (B), in which we see that the average particle current exhibits asymmetric behavior with respect to $E_d$.\\
\indent Finally, the third column shows the effect of a single and a double barrier when $\mu=0$ is fixed.
Let us begin with a single Schottky barrier on the right [Fig.~\ref{fig:ratchet}(c1)]: The electron- and hole-excitations 
can tunnel out to the right only if their energy is higher than the barrier. Since holes are more blocked than electrons by the low energy filter whatever the value of $E_s^R$, the current is expected to flow rightward ($I^N_L>0$). Moreover, recalling that the tunneling rates between the localized states and the electronic reservoirs $L$ and $R$ are given by Eq.~\eqref{eq_rates1}, we infer that the dominant contributions to the current come from energy excitations roughly within $k_B T$ around $\mu$.
As a consequence the net electric current is expected to increase until the barrier height $E_s \simeq k_BT$ above $\mu$: This is confirmed by looking at the corresponding current plot (C).
Fig.~\ref{fig:ratchet}(c2) illustrates the double barrier case, focusing on the situation $E_s^R= -E_s^L\equiv E_s$.  
The high energy filter acts on the left way out as the low energy filter acts on the right way out, upon inverting the role of electrons and holes.
Since holes flowing leftward are equivalent to electrons going rightward, this results in an enhanced ratchet effect and thus a larger current, as shown in Fig.~\ref{fig:ratchet}(C). 
As in the single barrier case, the maximum current is expected and indeed found at $E_s\simeq k_B T$.
The double barrier configuration ensures high performances, but is of difficult implementation.
In Fig.~\ref{fig:ratchet}(c3) we thus discuss the hybrid case, with an energy filter on the left, e.g. an embedded quantum dot close 
to the interface, and a single Schottky barrier on the right.  Raising the barrier height increases the current, 
but now the filter position ($E_d$) plays a role in determining its value.  
The $E_d=0$ case is similar to the single barrier one, because we have assumed $\Gamma=\epsilon>k_BT$, i.e., 
almost all relevant excitations are within $\Gamma$, and hence coupled to the left reservoir as if there was no barrier at all.
By similar arguments, the results for the $E_d=-\epsilon$ case are closer to the double barrier one: in this case the states 
above $E_d+\Gamma/2$ are blocked as if there was a low energy filter.\\
\indent The current plots in Fig.~\ref{fig:ratchet} show the ratchet effect to be much less pronounced for asymmetric tunnel contacts $\gamma_{eL}\neq\gamma_{eR}$.  For this reason, when discussing the device performance we will focus on the other cases only.
\section{Device performance for energy harvesting and cooling}
\label{sec:results}
\subsection{Non local thermopower, figure of merit and power factor}
We define a \textit{non local} thermopower\cite{Mazza2014} quantifying the voltage ($\delta\mu/e$, $e<0$ electron charge) 
between the electronic reservoirs $L$ and $R$ due to the temperature difference ($\delta T_P$) with the phonon bath $P$, in the absence of a temperature bias between the two electrodes ($\delta T=0$):
\be
S=-\left.\frac{\delta\mu/e}{\delta T_P}\right|_{I^N_L=0}=\frac{1}{e\,T}\frac{L_{13}}{L_{11}}.
\label{eq:Snloc}
\ee
Similarly, the non local electronic\footnote{We refer to the ``electronic'' figure of merit $ZT$ to distinguish it from the ``full'' figure of merit $\overline{ZT}$, which would include also the phononic contributions to the thermal conductance, here neglected.} figure of merit $ZT$ and power factor $Q$ read
\begin{align}
ZT& = \frac{G_l S^2}{\Xi_l^\text{(P)}}T =\frac{L_{13}^2}{L_{11}L_{33}-L_{13}^2},\cr
Q& = G_l S^2 = \frac{1}{T^3}\frac{L_{13}^2}{L_{11}},
\label{eq:ZTQnloc}
\end{align}
where $G_l=\left.[e I^N_L/(\delta\mu/e)]\right|_{\delta T,\delta T_P=0}=e^2 L_{11}/T$ and $\Xi_l^\text{(P)}=\left.[I^Q_P/\delta T_P]\right|_{I^N_L=0}=(L_{11}L_{33}-L_{13}^2)/(T^2 L_{11})$ are local electrical and (electronic) thermal conductances.
\footnote{These coefficients are all special instances of the more general ones discussed in Ref.~\onlinecite{Mazza2014}.}
Recall that the figure of merit $ZT$ is enough to fully characterize the performance 
of a thermoelectric device in the linear response regime:\cite{Benenti2013,Callen1985} the maximum efficiency and the efficiency at maximum power (energy harvesting) and the coefficient of performance (cooling) can be expressed in terms 
of $ZT$ and the Carnot efficiency $\eta_C$. 
The power factor $Q$ is instead a measure of the maximum output power that can be delivered by the device when it works as a thermal machine.
\begin{figure}
     \begin{center}
 	    \subfigure{%
            \includegraphics[width=\columnwidth]{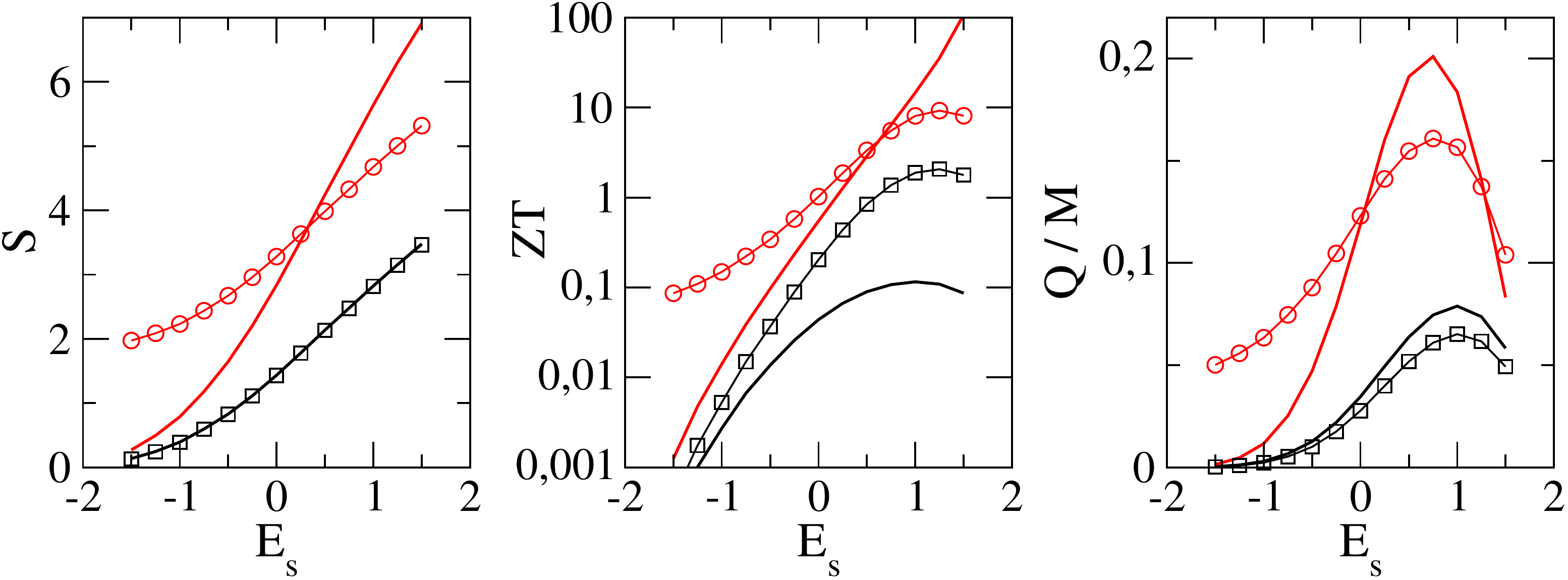}
        }\\[-0.2cm]
        \subfigure{%
           \includegraphics[width=\columnwidth]{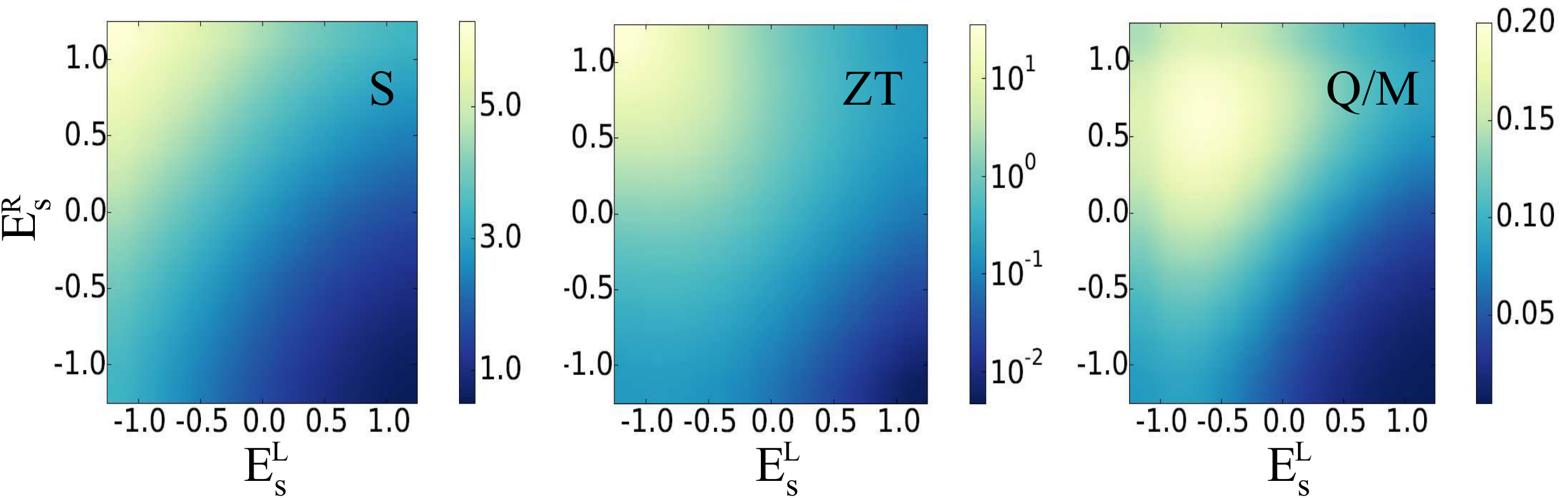}
        }\\[-0.5cm]
    \end{center}
    \caption{(Color online) (Top) Non local thermopower $S$ (in units of $k_B/e$), electronic figure of merit $ZT$, and power factor $Q$ (in units of $k_B^2/\hbar$) in the single barrier (black line), double barrier (red line) and hybrid (with $E_d=0$ ({\tiny{\color{black}$\square$}}) and $E_d=-\epsilon$ ({\large{\color{red}$\circ$}})) configurations. Data are plotted as functions of the (right) barrier height $E_s^R\equiv E_s$ (in units of $\epsilon$). For the double barrier case, $E_s^R = -E_s^L$. For the hybrid one, $\Gamma=\epsilon$. (Bottom) $S$, $ZT$, and $Q/M$ in the double barrier configuration as functions of the left and right barrier heights $E_s^L$ and $E_s^R$ (in units of $\epsilon$). In all panels, $\gamma_e=\epsilon/\hbar$ and $\mu=0$.}
   \label{fig:SZTP}
\end{figure}
In Sec.~\ref{sec:conversion} three situations were discussed: the single barrier, the double barrier, and the hybrid case.
Fig.~\ref{fig:SZTP}, top panels, show the non local coefficients $S$, $ZT$ and $Q$ in the three configurations, as functions of the right Schottky barrier height $E_s^R\equiv E_s$ (a comparison with the local coefficients is provided in Appendix~\ref{app_local}).
For the double barrier case, we have assumed $E_s^R=-E_s^L$, whereas for the hybrid case different choices 
for the center $E_d$ of the left filter are specified in the figure.
In all cases, the non local thermopower increases monotonically with $E_s$. 
Interestingly, the double barrier curve is exactly twice the single barrier one\footnote{Switching from the single barrier configuration to the double barrier configuration, the Onsager coefficient $L_{13}$ is increased while $L_{11}$ is decreased, in such a way that the ratio $L_{13}/L_{11}$ is exactly doubled for any value of $E_s$.}.
This can be understood by noticing that the two barriers at the right and left interfaces play the same role
in filtering electrons and holes, respectively, and their effects add up.  
This idealized configuration offers the largest thermopower at large $E_s$.
As a possible practical realization, we
consider the hybrid configuration with a large opening $\Gamma$ of the dot and a proper tuning of the level $E_d$.
With $E_d=-\epsilon$, we find that the thermopower in the hybrid case reaches values (red circles in the top left panel)
significantly larger than the single barrier one.  However, even if at small $E_s\lesssim 0$ this configuration is better 
than the ideal double barrier, it is not as performant at higher $E_s$.
On the other hand, we note that the $E_d = 0$ case is equivalent to the single barrier one [see Fig.~\ref{fig:ratchet}(b2) and related discussion].\\
\indent The non-local electronic figure of merit increases very rapidly with $E_s$. In particular, for the hybrid case it reaches
substantial values, up to $ZT\simeq 10$, halfway between the single barrier case ($ZT\simeq 0.1$) and the (ideal) double barrier one ($ZT\simeq 10^2$). Remarkably, the power factor $Q$ shows a similar behavior as $ZT$: It increases with the Schottky barrier height at least up to $E_s\simeq k_BT$, and then starts decreasing.  
Just as for the particle current discussed previously and shown in Fig.~\ref{fig:ratchet}(C),
this is because the NW states better coupled to the reservoirs are those within $k_BT$ of $\mu$. Note that for the hybrid case, the value of $E_s$ that maximizes $ZT$ is (almost) equal to the one maximizing $Q$: high $ZT\approx 10$ can be reached together with a good (scalable) power factor.\\
\indent For completeness, we have also plotted in the bottom panels of Fig.~\ref{fig:SZTP} the non local transport coefficients $S$, $ZT$ and $Q$ in a general double barrier configuration, as functions of the barriers heights $E_s^L$ and $E_s^R$.
Given our system, a symmetry with respect to the axis $E_s^R=-E_s^L$ arises.
Quite more remarkably, we observe a wide range of values
of $E_s^R$ and $E_s^L$ for which the electronic figure of merit and the power factor are \textit{simultaneously} large.
In particular, in the regime where $Q/M$ is maximum (around $E_s^R=k_BT=0.5\,\epsilon=-E_s^L$) $ZT$ can reach values up to $\sim 10$. 

\subsection{Energy harvesting and cooling}

\begin{figure}[b!]
    \includegraphics[clip,keepaspectratio,width=\columnwidth]{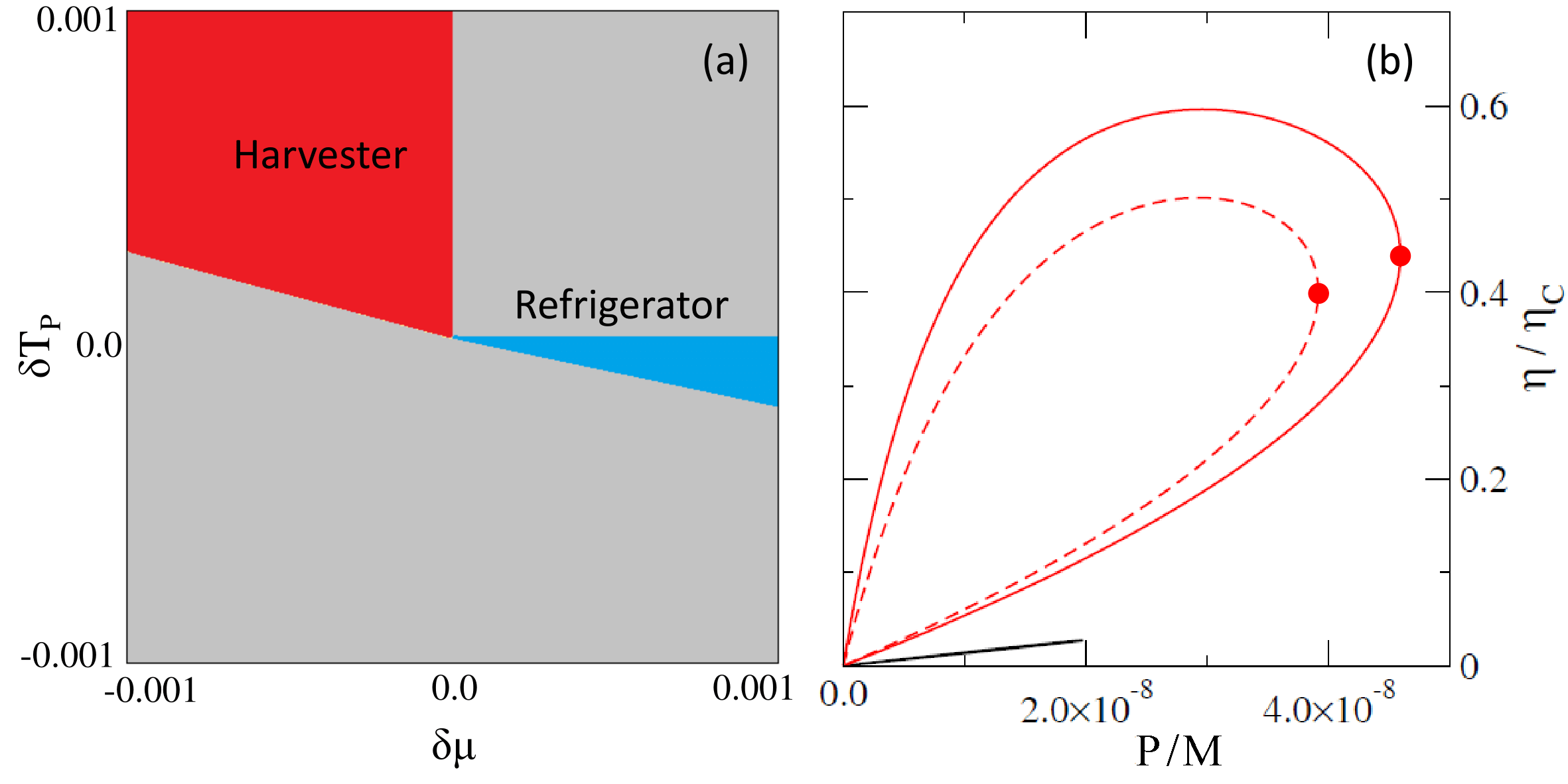}
    \caption{(Color online) (a) Phase diagram showing for which values of $\delta\mu$ (in units of $\epsilon$) and $\delta T_P$ (in units of $\epsilon/k_B$) the NW-based ratchet operates as an energy harvester or as a phonon bath refrigerator, given that $\delta T= 0$. The plot is drawn for the hybrid configuration (with $E_d=-\epsilon$, $\Gamma=\epsilon$, and $E_s^R=\epsilon$), the single and double barrier ones being qualitatively similar. (b) Heat-to-work conversion efficiency $\eta$ (normalized to $\eta_C$) as function of the output power $P$ (in units of $\epsilon^2/\hbar$) when $\delta\mu$ is varied, at fixed $\delta T_P=10^{-3}\epsilon/k_B$. The three loops correspond to the single barrier ($E_s^R=\epsilon$, black line), double barrier ($E^R_s=-E_s^L=\epsilon$, red line), and hybrid [same as in (a), red dashed line] cases. The red dots highlight the values of the efficiency at maximum power $\eta(P_{\rm max})$. In both panels, $\mu=0$ and $\gamma_e=\epsilon/\hbar$.}
   \label{fig:efficiency}
\end{figure}   

In order to harvest power from the device, we apply a bias $\delta\mu<0$ against the particle current.
We focus on the regime where electrical power $P=-\delta\mu I^N_L>0$ can be generated from waste heat ($I^Q_P>0$)
from the hot substrate ($\delta T_P>0$) under the condition of isothermal electronic reservoirs ($\delta T=0$).
This is a particular region of the phase diagram shown in Fig.~\ref{fig:efficiency}(a),  
delimited by\cite{Benenti2013,Nakpathomkun2010} $\delta\mu_\text{stop}<\delta\mu<0$ where $\delta\mu_\text{stop}\equiv -eS\delta T_P$ is the critical value at which the output power vanishes.
By further decreasing $\delta\mu$, $P$ becomes negative and hence the harvester useless.
Note that $\delta\mu_\text{stop}$ is proportional to the thermopower $S$ plotted in the top left panel of Fig.~\ref{fig:SZTP}. This implies in particular that the working regime range is broader in the double barrier case than in the single barrier one.\\
\indent The heat-to-work conversion efficiency $\eta$ of the ratchet in this harvesting regime is simply given by\cite{Benenti2013} $\eta=-\delta\mu I^N_L/I^Q_P$ because here, and in all cases considered in this paper, $I^Q_R<0$ and $I^Q_L<0$. In Fig.~\ref{fig:efficiency}(b),    $\eta$ is plotted  as function of the output power $P=-\delta\mu I^N_L$, upon varying $\delta\mu$ from $0$ to $\delta\mu_\text{stop}$, keeping $\delta T_P$ fixed.
Whilst in the single barrier setup both efficiencies and output power are small, the double barrier and hybrid configurations allow to extract a much larger power with an efficiency up to $60\%$ of the Carnot limit $\eta_C\equiv 1- T/T_P\simeq\delta T_P/T$.
Notice also that the efficiency at maximum output power $\eta(P_\text{max})$ -- highlighted by the red dots in Fig.~\ref{fig:efficiency}(b) -- can reach values close to the Curzon-Ahlborn limit\cite{CurzonAhlborn1975,Benenti2013} $\eta_{CA}\simeq\eta_C/2$.  
Concerning $P_\text{max}$, we find that a value of $P/M\approx 4.10^{-8}\,\epsilon^2/\hbar$ in Fig.~\ref{fig:efficiency}(b), obtained with $\delta T_P= 10^{-3}\,\epsilon/k_B$ and $\gamma_{e}=\gamma_{ep}=\epsilon/\hbar$, corresponds to $P/M\approx 10^{-15}\,\mathrm{W}$ assuming $\epsilon/k_B\approx 100\,\mathrm{K}$ (hence $\delta T_P\approx 0.1\,\mathrm{K}$, and $\gamma_{e}=\gamma_{ep}\approx 1.3\times 10^{13}\,\mathrm{s}^{-1}$). Consequently, for an array of $M=10^6$ NWs and a larger temperature bias $\delta T_P\approx 10\,\mathrm{K}$, a maximum output power of the order of $P_{\text{max}}\approx 10\,\mu\mathrm{W}$ can be envisaged\footnote{It is also interesting to give an estimation of the reachable output power density. It depends on the geometric parameters of the NW array. They must be chosen so as to be consistent with the experimental constraints and with the 1D model used in this work (typically, the NW diameter needs to be smaller than the Mott hopping length\cite{Bosisio20142} $L_M\approx\sqrt{\xi/(2\nu k_BT)}\approx 4a$). Also, since the reported results are mostly independent of the NW length $L$ as long as $L\gg L_M$ is long enough to be in the Mott hopping regime, choosing a short $L$ is preferable. By considering for instance a NW diameter of $20\,\mathrm{nm}$ and a packing density of $20\%$, $M=10^5$ NWs can be stacked in parallel along a $1\,\mathrm{cm}$ wide chip, yielding $P_{\text{max}}\approx 1\,\mu\mathrm{W}$ (keeping the same model parameters as right above).  Taking $1\,\mu\mathrm{m}$-long NWs, this gives an output power density of the order of $10\,\mathrm{mW/cm}^2$.}. Recall however that this is an "electronic" estimate.  The full $\eta$ will be decreased by parasitic phonon contributions here neglected.\\
\indent Beside harvesting power, the device could also be used as a refrigerator of the phonon bath.\cite{Jiang2012}  
In this case $\delta T_P<0$ and $\delta\mu>0$: an electrical power $\delta\mu I^N_L>0$ is invested to extract heat $I^Q_P>0$ 
from the (cold) substrate. The refrigerator working range is $\delta\mu>\delta\mu_{stop}^{(r)}\equiv-L_{33}/(L_{13}T)\delta T_P$. At the critical value $\delta\mu_{stop}^{(r)}$ the heat current from the phonon bath vanishes, $I^Q_P=0$.
Fig.~\ref{fig:efficiency}(a) shows that the refrigerator working region (blue) is smaller than the harvesting one (red).
Besides, the cooling efficiency of the ratchet is the coefficient of performance $\eta^{(r)}=I^Q_P/\delta\mu I^N_L$, 
characterized by the same electronic figure of merit $ZT$ as in the energy harvesting case.
Hence, though the double-barrier setup is once again the ideal one, the hybrid configuration allows to reach $ZT\simeq 10$,
making the NW-based ratchet a potentially high-performance cooler.

\section{Conclusion}
\label{ccl}
We have discussed the possible realization of a semiconductor NW-based ratchet for thermoelectric applications, 
operating in the activated hopping regime.  We have shown how to exploit spatial symmetry breaking at the contacts 
for the generation of a finite electric current through the NWs, and analyzed several ways in which this could be achieved.
In particular the ``hybrid configuration'', which could be implemented experimentally by embedding a quantum dot close to one contact 
and fabricating a Schottky barrier at the other one, can achieve {\textit{simultaneously}} substantial efficiency, 
with an electronic figure of merit $ZT\sim 10$, and large (scalable) output power $Q$.
A more realistic estimate of the \textit{full} figure of merit $\overline{ZT}$ would need to evaluate also the parasitic (phononic) contributions to the heat conductance\cite{Jiang2012} between the two electrodes and the substrate\footnote{Additional heat exchanges between the substrate phonons and the electrodes could take place via direct contact between them, or via phonon-mediated processes involving NWs phonons.}.
These would reduce the device performance, but their effect can be limited by suitably engineering the geometry of the setup\cite{Venkatasubramanian2000,Heron2010,Yu2010,He2012}. All these considerations put forward the proposed NW-based ratchet as a simple, reliable and high-performance thermoelectric setup, offering opportunities both for energy harvesting and for cooling.\\
\indent Future developments of this work may concern a time dependent control of the generated current. 
For instance, by acting with a time-varying gate potential on the energy filter level $E_d$ [see Fig.~\ref{fig:ratchet}(b1), (b2), (b3), and (c3)], one could arbitrarily tune the sign of $I^N_L$, thus exploiting the heat coming from the substrate to generate AC currents. More generally, the possibility of exploiting (further) ratchet effects due to time-dependent drivings
could be explored.\cite{Denisov2014}

\acknowledgments
We thank S. Roddaro for fruitful comments and suggestions, and the STherQO members for inspiring discussions.
The work of  R.B. has been supported by MIUR-FIRB2013 -- Project Coca (Grant No.~RBFR1379UX).
C.G. acknowledges financial support from the Deutsche Forschungsgemeinschaft through SFB 689.
C.G, G.F. and J.-L.P. acknowledge CEA for its support within the DSM-Energy Program (Project No. E112-7-Meso-Therm-DSM).

\appendix
\section{Temperature effects}
\label{app_T}

\begin{figure}
    \includegraphics[clip,keepaspectratio,width=0.75\columnwidth]{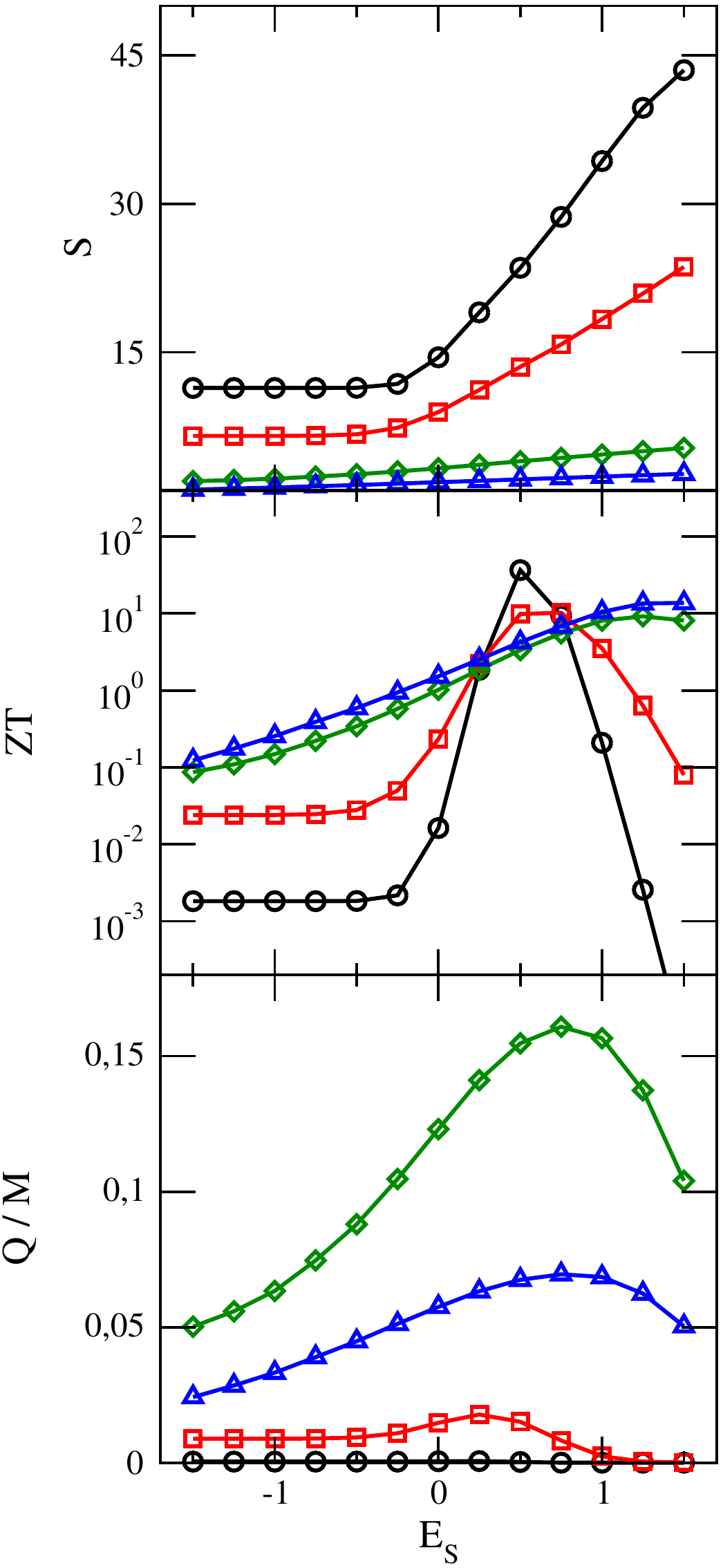}
    \caption{(Color online) Non local thermopower $S$ (in units of $k_B/e$), electronic figure of merit $ZT$ and power factor $Q$ (in units of $k_B^2/\hbar$) for various temperatures [$k_BT/\epsilon=0.05$ ({\large{\color{black}$\circ$}}), $0.1$ ({\tiny{\color{red}$\square$}}), $0.5$ ({\normalsize{\color{blue}$\diamond$}}) and $1$ ({\tiny{\color{green}$\triangle$}})]. Data are plotted for the hybrid configuration as functions of the (right) barrier height $E_s^R\equiv E_s$ (in units of $\epsilon$). Other parameters are $E_d=-\epsilon$, $\Gamma=\epsilon$, $\gamma_e=\epsilon/\hbar$, and $\mu=0$.}
   \label{fig:SZTP_variousT}
\end{figure}

In this section we estimate the results dependence on the temperature.
This issue was addressed in a previous work\cite{BosisioGorini2015} for a similar system under different conditions.
The (non local) coefficients $S$, $ZT$ and $Q$ are plotted in Fig.~\ref{fig:SZTP_variousT} for different $k_BT$'s, in the hybrid configuration. The thermopower reaches higher values at small temperatures. However, in this regime the electrical conductance $G_l$ is very small\cite{Bosisio20142}, and so is also the power factor $Q=G_l S^2$. 
This is evidence of the fact that the thermal energy $k_BT$ establishes how easy it is for a localized electron to hop toward another localized state in the (activated) hopping regime: If $k_BT$ is too small, the electrical conductance vanishes exponentially, reducing the power factor drastically.
Furthermore, it is also known\cite{Bosisio20142} that increasing the temperature too much reduces $G_l$ after some point, when all terms $I_{ij}^{(k)}$ and $I_{i\alpha}^{(k)}$, for each couple $(i,j)$, $(i,\alpha)$ and NW $k$, tend to vanish, irrespective of the degree of left-right asymmetry.
In the end, the best compromise for the power factor is found for an intermediate temperature $k_BT\simeq 0.5\epsilon$.
Concerning the electronic figure of merit $ZT$, its behavior with $T$ much depends on the right barrier height $E_s$. It reaches its highest value for $E_s\approx 0.5\,\epsilon$ at low temperatures, but at this point the smallness of the power factor limits the device performance. Nevertheless, a good compromise can be found between efficiency and output power in the temperature range $k_BT\approx 0.1-1\, \epsilon$ with a correct adjustment of $E_s$.

\section{Local vs Non local transport coefficients}
\label{app_local}

\begin{figure}
    \includegraphics[clip,keepaspectratio,width=0.75\columnwidth]{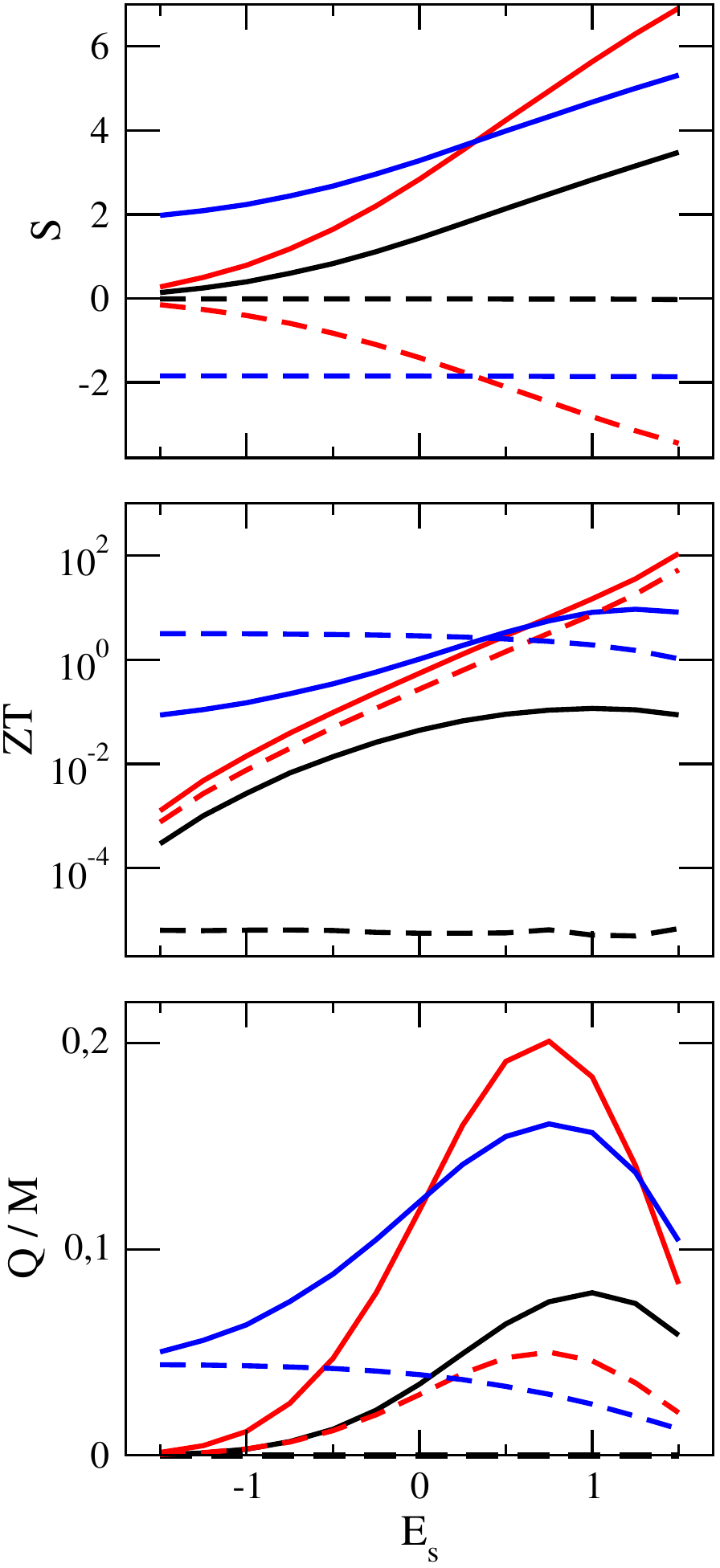}
    \caption{(Color online) Non local (full lines) versus local (dashed lines) transport coefficients, shown for the single barrier (black), double barrier (red) and hybrid (blue) configurations. Data are plotted as functions of the (right) barrier height $E_s^R\equiv E_s$ (in units of $\epsilon$). For the double barrier case, $E_s^R = -E_s^L$. For the hybrid one, $E_d=-\epsilon$ and $\Gamma=\epsilon$. The thermopowers and power factors are given in units of $k_B/e$ and $k_B^2/\hbar$ respectively. In all panels, $\gamma_e=\epsilon/\hbar$ and $\mu=0$.}
   \label{fig:SZTP_locnonloc}
\end{figure}

Our NW-based ratchet, in the configurations considered, boasts non-local transport coefficients 
typically larger than the local ones.  The latter are defined for $\delta T_P=0$ as:
\begin{align}
S_l&=-\left.\frac{\delta\mu}{e\delta T}\right|_{I^N_L=0}=\frac{L_{12}}{e\,TL_{11}},\cr
ZT_l& = \frac{G_l S_l^2}{\Xi_l^\text{(L)}}T =\frac{L_{12}^2}{L_{11}L_{22}-L_{12}^2},\cr
Q_l& = G_l S_l^2 = \frac{1}{T^3}\frac{L_{12}^2}{L_{11}},
\label{eq:SZTQnloc}
\end{align}
where $G_l=\left.[eI^N_L/(\delta\mu/e)]\right|_{\delta T,\delta T_P=0}=e^2 L_{11}/T$ and $\Xi_l^\text{(L)}=\left.[I^Q_L/\delta T]\right|_{I^N_L=0}=(L_{11}L_{22}-L_{12}^2)/(T^2 L_{11})$ are local electric and (electronic) thermal conductances\cite{Mazza2014}.
In Fig.~\ref{fig:SZTP_locnonloc} we show the local (dashed lines) and non local (full lines) transport coefficients for the single barrier, double barrier and hybrid configurations. In all the cases, the non local coefficients can reach larger values with respect to the corresponding local ones. Notice however that the local coefficients can be enhanced by probing band edge transport,
\cite{Bosisio20142,BosisioGorini2015} which is an alternative route to the symmetry-breaking one followed here.

\bibliography{Ratchet}

\begin{thebibliography}{74}%
\makeatletter
\providecommand \@ifxundefined [1]{%
 \@ifx{#1\undefined}
}%
\providecommand \@ifnum [1]{%
 \ifnum #1\expandafter \@firstoftwo
 \else \expandafter \@secondoftwo
 \fi
}%
\providecommand \@ifx [1]{%
 \ifx #1\expandafter \@firstoftwo
 \else \expandafter \@secondoftwo
 \fi
}%
\providecommand \natexlab [1]{#1}%
\providecommand \enquote  [1]{``#1''}%
\providecommand \bibnamefont  [1]{#1}%
\providecommand \bibfnamefont [1]{#1}%
\providecommand \citenamefont [1]{#1}%
\providecommand \href@noop [0]{\@secondoftwo}%
\providecommand \href [0]{\begingroup \@sanitize@url \@href}%
\providecommand \@href[1]{\@@startlink{#1}\@@href}%
\providecommand \@@href[1]{\endgroup#1\@@endlink}%
\providecommand \@sanitize@url [0]{\catcode `\\12\catcode `\$12\catcode
  `\&12\catcode `\#12\catcode `\^12\catcode `\_12\catcode `\%12\relax}%
\providecommand \@@startlink[1]{}%
\providecommand \@@endlink[0]{}%
\providecommand \url  [0]{\begingroup\@sanitize@url \@url }%
\providecommand \@url [1]{\endgroup\@href {#1}{\urlprefix }}%
\providecommand \urlprefix  [0]{URL }%
\providecommand \Eprint [0]{\href }%
\providecommand \doibase [0]{http://dx.doi.org/}%
\providecommand \selectlanguage [0]{\@gobble}%
\providecommand \bibinfo  [0]{\@secondoftwo}%
\providecommand \bibfield  [0]{\@secondoftwo}%
\providecommand \translation [1]{[#1]}%
\providecommand \BibitemOpen [0]{}%
\providecommand \bibitemStop [0]{}%
\providecommand \bibitemNoStop [0]{.\EOS\space}%
\providecommand \EOS [0]{\spacefactor3000\relax}%
\providecommand \BibitemShut  [1]{\csname bibitem#1\endcsname}%
\let\auto@bib@innerbib\@empty
%</preamble>
\bibitem [{\citenamefont {Ioffe}(1957)}]{Ioffe1957}%
  \BibitemOpen
  \bibfield  {author} {\bibinfo {author} {\bibfnamefont {A.~F.}\ \bibnamefont
  {Ioffe}},\ }\href@noop {} {\emph {\bibinfo {title} {Semiconductor
  Thermoelements and Thermoelectric Cooling}}}\ (\bibinfo  {publisher}
  {Infosearch},\ \bibinfo {year} {1957})\BibitemShut {NoStop}%
\bibitem [{\citenamefont {Hicks}\ and\ \citenamefont
  {Dresselhaus}(1993{\natexlab{a}})}]{Hicks1993}%
  \BibitemOpen
  \bibfield  {author} {\bibinfo {author} {\bibfnamefont {L.~D.}\ \bibnamefont
  {Hicks}}\ and\ \bibinfo {author} {\bibfnamefont {M.~S.}\ \bibnamefont
  {Dresselhaus}},\ }\href {\doibase 10.1103/PhysRevB.47.12727} {\bibfield
  {journal} {\bibinfo  {journal} {Phys. Rev. B}\ }\textbf {\bibinfo {volume}
  {47}},\ \bibinfo {pages} {12727} (\bibinfo {year}
  {1993}{\natexlab{a}})}\BibitemShut {NoStop}%
\bibitem [{\citenamefont {Hicks}\ and\ \citenamefont
  {Dresselhaus}(1993{\natexlab{b}})}]{Hicks1993bis}%
  \BibitemOpen
  \bibfield  {author} {\bibinfo {author} {\bibfnamefont {L.~D.}\ \bibnamefont
  {Hicks}}\ and\ \bibinfo {author} {\bibfnamefont {M.~S.}\ \bibnamefont
  {Dresselhaus}},\ }\href {\doibase 10.1103/PhysRevB.47.16631} {\bibfield
  {journal} {\bibinfo  {journal} {Phys. Rev. B}\ }\textbf {\bibinfo {volume}
  {47}},\ \bibinfo {pages} {16631} (\bibinfo {year}
  {1993}{\natexlab{b}})}\BibitemShut {NoStop}%
\bibitem [{\citenamefont {Saito}\ \emph {et~al.}(2011)\citenamefont {Saito},
  \citenamefont {Benenti}, \citenamefont {Casati},\ and\ \citenamefont
  {Prosen}}]{Saito2011}%
  \BibitemOpen
  \bibfield  {author} {\bibinfo {author} {\bibfnamefont {K.}~\bibnamefont
  {Saito}}, \bibinfo {author} {\bibfnamefont {G.}~\bibnamefont {Benenti}},
  \bibinfo {author} {\bibfnamefont {G.}~\bibnamefont {Casati}}, \ and\ \bibinfo
  {author} {\bibfnamefont {T.}~\bibnamefont {Prosen}},\ }\href {\doibase
  10.1103/PhysRevB.84.201306} {\bibfield  {journal} {\bibinfo  {journal} {Phys.
  Rev. B}\ }\textbf {\bibinfo {volume} {84}},\ \bibinfo {pages} {201306}
  (\bibinfo {year} {2011})}\BibitemShut {NoStop}%
\bibitem [{\citenamefont {S\'anchez}\ and\ \citenamefont
  {Serra}(2011)}]{SanchezDavid2011}%
  \BibitemOpen
  \bibfield  {author} {\bibinfo {author} {\bibfnamefont {D.}~\bibnamefont
  {S\'anchez}}\ and\ \bibinfo {author} {\bibfnamefont {L.}~\bibnamefont
  {Serra}},\ }\href {\doibase 10.1103/PhysRevB.84.201307} {\bibfield  {journal}
  {\bibinfo  {journal} {Phys. Rev. B}\ }\textbf {\bibinfo {volume} {84}},\
  \bibinfo {pages} {201307} (\bibinfo {year} {2011})}\BibitemShut {NoStop}%
\bibitem [{\citenamefont {Horvat}\ \emph {et~al.}(2012)\citenamefont {Horvat},
  \citenamefont {Prosen}, \citenamefont {Benenti},\ and\ \citenamefont
  {Casati}}]{Horvat2012}%
  \BibitemOpen
  \bibfield  {author} {\bibinfo {author} {\bibfnamefont {M.}~\bibnamefont
  {Horvat}}, \bibinfo {author} {\bibfnamefont {T.}~\bibnamefont {Prosen}},
  \bibinfo {author} {\bibfnamefont {G.}~\bibnamefont {Benenti}}, \ and\
  \bibinfo {author} {\bibfnamefont {G.}~\bibnamefont {Casati}},\ }\href
  {\doibase 10.1103/PhysRevE.86.052102} {\bibfield  {journal} {\bibinfo
  {journal} {Phys. Rev. E}\ }\textbf {\bibinfo {volume} {86}},\ \bibinfo
  {pages} {052102} (\bibinfo {year} {2012})}\BibitemShut {NoStop}%
\bibitem [{\citenamefont {Balachandran}\ \emph {et~al.}(2013)\citenamefont
  {Balachandran}, \citenamefont {Benenti},\ and\ \citenamefont
  {Casati}}]{Balachandran2013}%
  \BibitemOpen
  \bibfield  {author} {\bibinfo {author} {\bibfnamefont {V.}~\bibnamefont
  {Balachandran}}, \bibinfo {author} {\bibfnamefont {G.}~\bibnamefont
  {Benenti}}, \ and\ \bibinfo {author} {\bibfnamefont {G.}~\bibnamefont
  {Casati}},\ }\href {\doibase 10.1103/PhysRevB.87.165419} {\bibfield
  {journal} {\bibinfo  {journal} {Phys. Rev. B}\ }\textbf {\bibinfo {volume}
  {87}},\ \bibinfo {pages} {165419} (\bibinfo {year} {2013})}\BibitemShut
  {NoStop}%
\bibitem [{\citenamefont {Brandner}\ \emph {et~al.}(2013)\citenamefont
  {Brandner}, \citenamefont {Saito},\ and\ \citenamefont
  {Seifert}}]{Brandner2013}%
  \BibitemOpen
  \bibfield  {author} {\bibinfo {author} {\bibfnamefont {K.}~\bibnamefont
  {Brandner}}, \bibinfo {author} {\bibfnamefont {K.}~\bibnamefont {Saito}}, \
  and\ \bibinfo {author} {\bibfnamefont {U.}~\bibnamefont {Seifert}},\ }\href
  {\doibase 10.1103/PhysRevLett.110.070603} {\bibfield  {journal} {\bibinfo
  {journal} {Phys. Rev. Lett.}\ }\textbf {\bibinfo {volume} {110}},\ \bibinfo
  {pages} {070603} (\bibinfo {year} {2013})}\BibitemShut {NoStop}%
\bibitem [{\citenamefont {Mazza}\ \emph {et~al.}(2014)\citenamefont {Mazza},
  \citenamefont {Bosisio}, \citenamefont {Benenti}, \citenamefont
  {Giovannetti}, \citenamefont {Fazio},\ and\ \citenamefont
  {Taddei}}]{Mazza2014}%
  \BibitemOpen
  \bibfield  {author} {\bibinfo {author} {\bibfnamefont {F.}~\bibnamefont
  {Mazza}}, \bibinfo {author} {\bibfnamefont {R.}~\bibnamefont {Bosisio}},
  \bibinfo {author} {\bibfnamefont {G.}~\bibnamefont {Benenti}}, \bibinfo
  {author} {\bibfnamefont {V.}~\bibnamefont {Giovannetti}}, \bibinfo {author}
  {\bibfnamefont {R.}~\bibnamefont {Fazio}}, \ and\ \bibinfo {author}
  {\bibfnamefont {F.}~\bibnamefont {Taddei}},\ }\href
  {http://stacks.iop.org/1367-2630/16/i=8/a=085001} {\bibfield  {journal}
  {\bibinfo  {journal} {New Journal of Physics}\ }\textbf {\bibinfo {volume}
  {16}},\ \bibinfo {pages} {085001} (\bibinfo {year} {2014})}\BibitemShut
  {NoStop}%
\bibitem [{\citenamefont {Bosisio}\ \emph
  {et~al.}(2015{\natexlab{a}})\citenamefont {Bosisio}, \citenamefont
  {Valentini}, \citenamefont {Mazza}, \citenamefont {Benenti}, \citenamefont
  {Fazio}, \citenamefont {Giovannetti},\ and\ \citenamefont
  {Taddei}}]{Bosisio2015}%
  \BibitemOpen
  \bibfield  {author} {\bibinfo {author} {\bibfnamefont {R.}~\bibnamefont
  {Bosisio}}, \bibinfo {author} {\bibfnamefont {S.}~\bibnamefont {Valentini}},
  \bibinfo {author} {\bibfnamefont {F.}~\bibnamefont {Mazza}}, \bibinfo
  {author} {\bibfnamefont {G.}~\bibnamefont {Benenti}}, \bibinfo {author}
  {\bibfnamefont {R.}~\bibnamefont {Fazio}}, \bibinfo {author} {\bibfnamefont
  {V.}~\bibnamefont {Giovannetti}}, \ and\ \bibinfo {author} {\bibfnamefont
  {F.}~\bibnamefont {Taddei}},\ }\href {\doibase 10.1103/PhysRevB.91.205420}
  {\bibfield  {journal} {\bibinfo  {journal} {Phys. Rev. B}\ }\textbf {\bibinfo
  {volume} {91}},\ \bibinfo {pages} {205420} (\bibinfo {year}
  {2015}{\natexlab{a}})}\BibitemShut {NoStop}%
\bibitem [{\citenamefont {Whitney}(2013)}]{Whitney2013}%
  \BibitemOpen
  \bibfield  {author} {\bibinfo {author} {\bibfnamefont {R.~S.}\ \bibnamefont
  {Whitney}},\ }\href {\doibase 10.1103/PhysRevB.87.115404} {\bibfield
  {journal} {\bibinfo  {journal} {Phys. Rev. B}\ }\textbf {\bibinfo {volume}
  {87}},\ \bibinfo {pages} {115404} (\bibinfo {year} {2013})}\BibitemShut
  {NoStop}%
\bibitem [{\citenamefont {Machon}\ \emph {et~al.}(2013)\citenamefont {Machon},
  \citenamefont {Eschrig},\ and\ \citenamefont {Belzig}}]{Machon2013}%
  \BibitemOpen
  \bibfield  {author} {\bibinfo {author} {\bibfnamefont {P.}~\bibnamefont
  {Machon}}, \bibinfo {author} {\bibfnamefont {M.}~\bibnamefont {Eschrig}}, \
  and\ \bibinfo {author} {\bibfnamefont {W.}~\bibnamefont {Belzig}},\ }\href
  {\doibase 10.1103/PhysRevLett.110.047002} {\bibfield  {journal} {\bibinfo
  {journal} {Phys. Rev. Lett.}\ }\textbf {\bibinfo {volume} {110}},\ \bibinfo
  {pages} {047002} (\bibinfo {year} {2013})}\BibitemShut {NoStop}%
\bibitem [{\citenamefont {{Mazza}}\ \emph {et~al.}(2015)\citenamefont
  {{Mazza}}, \citenamefont {{Valentini}}, \citenamefont {{Bosisio}},
  \citenamefont {{Benenti}}, \citenamefont {{Giovannetti}}, \citenamefont
  {{Fazio}},\ and\ \citenamefont {{Taddei}}}]{Mazza2015}%
  \BibitemOpen
  \bibfield  {author} {\bibinfo {author} {\bibfnamefont {F.}~\bibnamefont
  {{Mazza}}}, \bibinfo {author} {\bibfnamefont {S.}~\bibnamefont
  {{Valentini}}}, \bibinfo {author} {\bibfnamefont {R.}~\bibnamefont
  {{Bosisio}}}, \bibinfo {author} {\bibfnamefont {G.}~\bibnamefont
  {{Benenti}}}, \bibinfo {author} {\bibfnamefont {V.}~\bibnamefont
  {{Giovannetti}}}, \bibinfo {author} {\bibfnamefont {R.}~\bibnamefont
  {{Fazio}}}, \ and\ \bibinfo {author} {\bibfnamefont {F.}~\bibnamefont
  {{Taddei}}},\ }\href
  {http://journals.aps.org/prb/abstract/10.1103/PhysRevB.91.245435} {\bibfield
  {journal} {\bibinfo  {journal} {Phys. Rev. B}\ }\textbf {\bibinfo {volume}
  {91}},\ \bibinfo {pages} {245435} (\bibinfo {year} {2015})}\BibitemShut
  {NoStop}%
\bibitem [{\citenamefont {Valentini}\ \emph {et~al.}(2015)\citenamefont
  {Valentini}, \citenamefont {Fazio}, \citenamefont {Giovannetti},\ and\
  \citenamefont {Taddei}}]{Valentini2015}%
  \BibitemOpen
  \bibfield  {author} {\bibinfo {author} {\bibfnamefont {S.}~\bibnamefont
  {Valentini}}, \bibinfo {author} {\bibfnamefont {R.}~\bibnamefont {Fazio}},
  \bibinfo {author} {\bibfnamefont {V.}~\bibnamefont {Giovannetti}}, \ and\
  \bibinfo {author} {\bibfnamefont {F.}~\bibnamefont {Taddei}},\ }\href
  {\doibase 10.1103/PhysRevB.91.045430} {\bibfield  {journal} {\bibinfo
  {journal} {Phys. Rev. B}\ }\textbf {\bibinfo {volume} {91}},\ \bibinfo
  {pages} {045430} (\bibinfo {year} {2015})}\BibitemShut {NoStop}%
\bibitem [{\citenamefont {S\'anchez}\ and\ \citenamefont
  {B\"uttiker}(2011)}]{Sanchez2011}%
  \BibitemOpen
  \bibfield  {author} {\bibinfo {author} {\bibfnamefont {R.}~\bibnamefont
  {S\'anchez}}\ and\ \bibinfo {author} {\bibfnamefont {M.}~\bibnamefont
  {B\"uttiker}},\ }\href {\doibase 10.1103/PhysRevB.83.085428} {\bibfield
  {journal} {\bibinfo  {journal} {Phys. Rev. B}\ }\textbf {\bibinfo {volume}
  {83}},\ \bibinfo {pages} {085428} (\bibinfo {year} {2011})}\BibitemShut
  {NoStop}%
\bibitem [{\citenamefont {Jordan}\ \emph {et~al.}(2013)\citenamefont {Jordan},
  \citenamefont {Sothmann}, \citenamefont {S\'anchez},\ and\ \citenamefont
  {B\"uttiker}}]{Jordan2013}%
  \BibitemOpen
  \bibfield  {author} {\bibinfo {author} {\bibfnamefont {A.~N.}\ \bibnamefont
  {Jordan}}, \bibinfo {author} {\bibfnamefont {B.}~\bibnamefont {Sothmann}},
  \bibinfo {author} {\bibfnamefont {R.}~\bibnamefont {S\'anchez}}, \ and\
  \bibinfo {author} {\bibfnamefont {M.}~\bibnamefont {B\"uttiker}},\ }\href
  {\doibase 10.1103/PhysRevB.87.075312} {\bibfield  {journal} {\bibinfo
  {journal} {Phys. Rev. B}\ }\textbf {\bibinfo {volume} {87}},\ \bibinfo
  {pages} {075312} (\bibinfo {year} {2013})}\BibitemShut {NoStop}%
\bibitem [{\citenamefont {Sothmann}\ \emph {et~al.}(2013)\citenamefont
  {Sothmann}, \citenamefont {S\'anchez}, \citenamefont {Jordan},\ and\
  \citenamefont {B\"uttiker}}]{Sothmann2013}%
  \BibitemOpen
  \bibfield  {author} {\bibinfo {author} {\bibfnamefont {B.}~\bibnamefont
  {Sothmann}}, \bibinfo {author} {\bibfnamefont {R.}~\bibnamefont {S\'anchez}},
  \bibinfo {author} {\bibfnamefont {A.~N.}\ \bibnamefont {Jordan}}, \ and\
  \bibinfo {author} {\bibfnamefont {M.}~\bibnamefont {B\"uttiker}},\ }\href
  {http://stacks.iop.org/1367-2630/15/i=9/a=095021} {\bibfield  {journal}
  {\bibinfo  {journal} {New J. Phys.}\ }\textbf {\bibinfo {volume} {15}},\
  \bibinfo {pages} {095021} (\bibinfo {year} {2013})}\BibitemShut {NoStop}%
\bibitem [{\citenamefont {Roche}\ \emph {et~al.}(2015)\citenamefont {Roche},
  \citenamefont {Roulleau}, \citenamefont {Jullien}, \citenamefont {Jompol},
  \citenamefont {Farrer}, \citenamefont {Ritchie},\ and\ \citenamefont
  {Glattli}}]{Roche2015}%
  \BibitemOpen
  \bibfield  {author} {\bibinfo {author} {\bibfnamefont {B.}~\bibnamefont
  {Roche}}, \bibinfo {author} {\bibfnamefont {P.}~\bibnamefont {Roulleau}},
  \bibinfo {author} {\bibfnamefont {T.}~\bibnamefont {Jullien}}, \bibinfo
  {author} {\bibfnamefont {Y.}~\bibnamefont {Jompol}}, \bibinfo {author}
  {\bibfnamefont {I.}~\bibnamefont {Farrer}}, \bibinfo {author} {\bibfnamefont
  {D.}~\bibnamefont {Ritchie}}, \ and\ \bibinfo {author} {\bibfnamefont
  {D.}~\bibnamefont {Glattli}},\ }\href {http://dx.doi.org/10.1038/ncomms7738}
  {\bibfield  {journal} {\bibinfo  {journal} {Nat. Commun.}\ }\textbf {\bibinfo
  {volume} {6}},\ \bibinfo {pages} {6738} (\bibinfo {year} {2015})}\BibitemShut
  {NoStop}%
\bibitem [{\citenamefont {Hartmann}\ \emph {et~al.}(2015)\citenamefont
  {Hartmann}, \citenamefont {Pfeffer}, \citenamefont {H\"ofling}, \citenamefont
  {Kamp},\ and\ \citenamefont {Worschech}}]{Hartmann2015}%
  \BibitemOpen
  \bibfield  {author} {\bibinfo {author} {\bibfnamefont {F.}~\bibnamefont
  {Hartmann}}, \bibinfo {author} {\bibfnamefont {P.}~\bibnamefont {Pfeffer}},
  \bibinfo {author} {\bibfnamefont {S.}~\bibnamefont {H\"ofling}}, \bibinfo
  {author} {\bibfnamefont {M.}~\bibnamefont {Kamp}}, \ and\ \bibinfo {author}
  {\bibfnamefont {L.}~\bibnamefont {Worschech}},\ }\href {\doibase
  10.1103/PhysRevLett.114.146805} {\bibfield  {journal} {\bibinfo  {journal}
  {Phys. Rev. Lett.}\ }\textbf {\bibinfo {volume} {114}},\ \bibinfo {pages}
  {146805} (\bibinfo {year} {2015})}\BibitemShut {NoStop}%
\bibitem [{\citenamefont {Thierschmann}\ \emph {et~al.}(2015)\citenamefont
  {Thierschmann}, \citenamefont {Arnold}, \citenamefont {Mitterm\"uller},
  \citenamefont {Maier}, \citenamefont {Heyn}, \citenamefont {Hansen},
  \citenamefont {Buhmann},\ and\ \citenamefont {Molenkamp}}]{Thierschmann2015}%
  \BibitemOpen
  \bibfield  {author} {\bibinfo {author} {\bibfnamefont {H.}~\bibnamefont
  {Thierschmann}}, \bibinfo {author} {\bibfnamefont {F.}~\bibnamefont
  {Arnold}}, \bibinfo {author} {\bibfnamefont {M.}~\bibnamefont
  {Mitterm\"uller}}, \bibinfo {author} {\bibfnamefont {L.}~\bibnamefont
  {Maier}}, \bibinfo {author} {\bibfnamefont {C.}~\bibnamefont {Heyn}},
  \bibinfo {author} {\bibfnamefont {W.}~\bibnamefont {Hansen}}, \bibinfo
  {author} {\bibfnamefont {H.}~\bibnamefont {Buhmann}}, \ and\ \bibinfo
  {author} {\bibfnamefont {L.~W.}\ \bibnamefont {Molenkamp}},\ }\href
  {http://stacks.iop.org/1367-2630/17/i=11/a=113003} {\bibfield  {journal}
  {\bibinfo  {journal} {New Journal of Physics}\ }\textbf {\bibinfo {volume}
  {17}},\ \bibinfo {pages} {113003} (\bibinfo {year} {2015})}\BibitemShut
  {NoStop}%
\bibitem [{\citenamefont {Hofer}\ and\ \citenamefont
  {Sothmann}(2015)}]{Hofer2015}%
  \BibitemOpen
  \bibfield  {author} {\bibinfo {author} {\bibfnamefont {P.~P.}\ \bibnamefont
  {Hofer}}\ and\ \bibinfo {author} {\bibfnamefont {B.}~\bibnamefont
  {Sothmann}},\ }\href {\doibase 10.1103/PhysRevB.91.195406} {\bibfield
  {journal} {\bibinfo  {journal} {Phys. Rev. B}\ }\textbf {\bibinfo {volume}
  {91}},\ \bibinfo {pages} {195406} (\bibinfo {year} {2015})}\BibitemShut
  {NoStop}%
\bibitem [{\citenamefont {S\'anchez}\ \emph {et~al.}(2015)\citenamefont
  {S\'anchez}, \citenamefont {Sothmann},\ and\ \citenamefont
  {Jordan}}]{Sanchez2015}%
  \BibitemOpen
  \bibfield  {author} {\bibinfo {author} {\bibfnamefont {R.}~\bibnamefont
  {S\'anchez}}, \bibinfo {author} {\bibfnamefont {B.}~\bibnamefont {Sothmann}},
  \ and\ \bibinfo {author} {\bibfnamefont {A.~N.}\ \bibnamefont {Jordan}},\
  }\href {\doibase 10.1103/PhysRevLett.114.146801} {\bibfield  {journal}
  {\bibinfo  {journal} {Phys. Rev. Lett.}\ }\textbf {\bibinfo {volume} {114}},\
  \bibinfo {pages} {146801} (\bibinfo {year} {2015})}\BibitemShut {NoStop}%
\bibitem [{\citenamefont {Rutten}\ \emph {et~al.}(2009)\citenamefont {Rutten},
  \citenamefont {Esposito},\ and\ \citenamefont {Cleuren}}]{Rutten2009}%
  \BibitemOpen
  \bibfield  {author} {\bibinfo {author} {\bibfnamefont {B.}~\bibnamefont
  {Rutten}}, \bibinfo {author} {\bibfnamefont {M.}~\bibnamefont {Esposito}}, \
  and\ \bibinfo {author} {\bibfnamefont {B.}~\bibnamefont {Cleuren}},\ }\href
  {\doibase 10.1103/PhysRevB.80.235122} {\bibfield  {journal} {\bibinfo
  {journal} {Phys. Rev. B}\ }\textbf {\bibinfo {volume} {80}},\ \bibinfo
  {pages} {235122} (\bibinfo {year} {2009})}\BibitemShut {NoStop}%
\bibitem [{\citenamefont {Ruokola}\ and\ \citenamefont
  {Ojanen}(2012)}]{Ruokola2012}%
  \BibitemOpen
  \bibfield  {author} {\bibinfo {author} {\bibfnamefont {T.}~\bibnamefont
  {Ruokola}}\ and\ \bibinfo {author} {\bibfnamefont {T.}~\bibnamefont
  {Ojanen}},\ }\href {\doibase 10.1103/PhysRevB.86.035454} {\bibfield
  {journal} {\bibinfo  {journal} {Phys. Rev. B}\ }\textbf {\bibinfo {volume}
  {86}},\ \bibinfo {pages} {035454} (\bibinfo {year} {2012})}\BibitemShut
  {NoStop}%
\bibitem [{\citenamefont {Bergenfeldt}\ \emph {et~al.}(2014)\citenamefont
  {Bergenfeldt}, \citenamefont {Samuelsson}, \citenamefont {Sothmann},
  \citenamefont {Flindt},\ and\ \citenamefont {B\"uttiker}}]{Bergenfeldt2014}%
  \BibitemOpen
  \bibfield  {author} {\bibinfo {author} {\bibfnamefont {C.}~\bibnamefont
  {Bergenfeldt}}, \bibinfo {author} {\bibfnamefont {P.}~\bibnamefont
  {Samuelsson}}, \bibinfo {author} {\bibfnamefont {B.}~\bibnamefont
  {Sothmann}}, \bibinfo {author} {\bibfnamefont {C.}~\bibnamefont {Flindt}}, \
  and\ \bibinfo {author} {\bibfnamefont {M.}~\bibnamefont {B\"uttiker}},\
  }\href {\doibase 10.1103/PhysRevLett.112.076803} {\bibfield  {journal}
  {\bibinfo  {journal} {Phys. Rev. Lett.}\ }\textbf {\bibinfo {volume} {112}},\
  \bibinfo {pages} {076803} (\bibinfo {year} {2014})}\BibitemShut {NoStop}%
\bibitem [{\citenamefont {Cleuren}\ \emph {et~al.}(2012)\citenamefont
  {Cleuren}, \citenamefont {Rutten},\ and\ \citenamefont {Van~den
  Broeck}}]{Cleuren2012}%
  \BibitemOpen
  \bibfield  {author} {\bibinfo {author} {\bibfnamefont {B.}~\bibnamefont
  {Cleuren}}, \bibinfo {author} {\bibfnamefont {B.}~\bibnamefont {Rutten}}, \
  and\ \bibinfo {author} {\bibfnamefont {C.}~\bibnamefont {Van~den Broeck}},\
  }\href@noop {} {\bibfield  {journal} {\bibinfo  {journal} {Phys. Rev. Lett.}\
  }\textbf {\bibinfo {volume} {108}},\ \bibinfo {pages} {120603} (\bibinfo
  {year} {2012})}\BibitemShut {NoStop}%
\bibitem [{\citenamefont {Mari}\ and\ \citenamefont {Eisert}(2012)}]{Mari2012}%
  \BibitemOpen
  \bibfield  {author} {\bibinfo {author} {\bibfnamefont {A.}~\bibnamefont
  {Mari}}\ and\ \bibinfo {author} {\bibfnamefont {J.}~\bibnamefont {Eisert}},\
  }\href@noop {} {\bibfield  {journal} {\bibinfo  {journal} {Phys. Rev. Lett.}\
  }\textbf {\bibinfo {volume} {108}},\ \bibinfo {pages} {120602} (\bibinfo
  {year} {2012})}\BibitemShut {NoStop}%
\bibitem [{\citenamefont {Entin-Wohlman}\ \emph {et~al.}(2010)\citenamefont
  {Entin-Wohlman}, \citenamefont {Imry},\ and\ \citenamefont
  {Aharony}}]{Entin2010}%
  \BibitemOpen
  \bibfield  {author} {\bibinfo {author} {\bibfnamefont {O.}~\bibnamefont
  {Entin-Wohlman}}, \bibinfo {author} {\bibfnamefont {Y.}~\bibnamefont {Imry}},
  \ and\ \bibinfo {author} {\bibfnamefont {A.}~\bibnamefont {Aharony}},\ }\href
  {\doibase 10.1103/PhysRevB.82.115314} {\bibfield  {journal} {\bibinfo
  {journal} {Phys. Rev. B}\ }\textbf {\bibinfo {volume} {82}},\ \bibinfo
  {pages} {115314} (\bibinfo {year} {2010})}\BibitemShut {NoStop}%
\bibitem [{\citenamefont {Jiang}\ \emph
  {et~al.}(2013{\natexlab{a}})\citenamefont {Jiang}, \citenamefont
  {Entin-Wohlman},\ and\ \citenamefont {Imry}}]{Jiang2013bis}%
  \BibitemOpen
  \bibfield  {author} {\bibinfo {author} {\bibfnamefont {J.-H.}\ \bibnamefont
  {Jiang}}, \bibinfo {author} {\bibfnamefont {O.}~\bibnamefont
  {Entin-Wohlman}}, \ and\ \bibinfo {author} {\bibfnamefont {Y.}~\bibnamefont
  {Imry}},\ }\href {http://stacks.iop.org/1367-2630/15/i=7/a=075021} {\bibfield
   {journal} {\bibinfo  {journal} {New Journal of Physics}\ }\textbf {\bibinfo
  {volume} {15}},\ \bibinfo {pages} {075021} (\bibinfo {year}
  {2013}{\natexlab{a}})}\BibitemShut {NoStop}%
\bibitem [{\citenamefont {Entin-Wohlman}\ \emph {et~al.}(2015)\citenamefont
  {Entin-Wohlman}, \citenamefont {Imry},\ and\ \citenamefont
  {Aharony}}]{Entin2015}%
  \BibitemOpen
  \bibfield  {author} {\bibinfo {author} {\bibfnamefont {O.}~\bibnamefont
  {Entin-Wohlman}}, \bibinfo {author} {\bibfnamefont {Y.}~\bibnamefont {Imry}},
  \ and\ \bibinfo {author} {\bibfnamefont {A.}~\bibnamefont {Aharony}},\ }\href
  {\doibase 10.1103/PhysRevB.91.054302} {\bibfield  {journal} {\bibinfo
  {journal} {Phys. Rev. B}\ }\textbf {\bibinfo {volume} {91}},\ \bibinfo
  {pages} {054302} (\bibinfo {year} {2015})}\BibitemShut {NoStop}%
\bibitem [{\citenamefont {Jiang}\ \emph {et~al.}(2012)\citenamefont {Jiang},
  \citenamefont {Entin-Wohlman},\ and\ \citenamefont {Imry}}]{Jiang2012}%
  \BibitemOpen
  \bibfield  {author} {\bibinfo {author} {\bibfnamefont {J.-H.}\ \bibnamefont
  {Jiang}}, \bibinfo {author} {\bibfnamefont {O.}~\bibnamefont
  {Entin-Wohlman}}, \ and\ \bibinfo {author} {\bibfnamefont {Y.}~\bibnamefont
  {Imry}},\ }\href@noop {} {\bibfield  {journal} {\bibinfo  {journal} {Phys.
  Rev. B}\ }\textbf {\bibinfo {volume} {85}},\ \bibinfo {pages} {075412}
  (\bibinfo {year} {2012})}\BibitemShut {NoStop}%
\bibitem [{\citenamefont {Jiang}\ \emph {et~al.}(2015)\citenamefont {Jiang},
  \citenamefont {Kulkarni}, \citenamefont {Segal},\ and\ \citenamefont
  {Imry}}]{Jiang2015}%
  \BibitemOpen
  \bibfield  {author} {\bibinfo {author} {\bibfnamefont {J.-H.}\ \bibnamefont
  {Jiang}}, \bibinfo {author} {\bibfnamefont {M.}~\bibnamefont {Kulkarni}},
  \bibinfo {author} {\bibfnamefont {D.}~\bibnamefont {Segal}}, \ and\ \bibinfo
  {author} {\bibfnamefont {Y.}~\bibnamefont {Imry}},\ }\href {\doibase
  10.1103/PhysRevB.92.045309} {\bibfield  {journal} {\bibinfo  {journal} {Phys.
  Rev. B}\ }\textbf {\bibinfo {volume} {92}},\ \bibinfo {pages} {045309}
  (\bibinfo {year} {2015})}\BibitemShut {NoStop}%
\bibitem [{\citenamefont {Bosisio}\ \emph
  {et~al.}(2014{\natexlab{a}})\citenamefont {Bosisio}, \citenamefont {Gorini},
  \citenamefont {Fleury},\ and\ \citenamefont {Pichard}}]{Bosisio20142}%
  \BibitemOpen
  \bibfield  {author} {\bibinfo {author} {\bibfnamefont {R.}~\bibnamefont
  {Bosisio}}, \bibinfo {author} {\bibfnamefont {C.}~\bibnamefont {Gorini}},
  \bibinfo {author} {\bibfnamefont {G.}~\bibnamefont {Fleury}}, \ and\ \bibinfo
  {author} {\bibfnamefont {J.-L.}\ \bibnamefont {Pichard}},\ }\href@noop {}
  {\bibfield  {journal} {\bibinfo  {journal} {New J. Phys.}\ }\textbf {\bibinfo
  {volume} {16}},\ \bibinfo {pages} {095005} (\bibinfo {year}
  {2014}{\natexlab{a}})}\BibitemShut {NoStop}%
\bibitem [{\citenamefont {Bosisio}\ \emph
  {et~al.}(2015{\natexlab{b}})\citenamefont {Bosisio}, \citenamefont {Gorini},
  \citenamefont {Fleury},\ and\ \citenamefont {Pichard}}]{BosisioGorini2015}%
  \BibitemOpen
  \bibfield  {author} {\bibinfo {author} {\bibfnamefont {R.}~\bibnamefont
  {Bosisio}}, \bibinfo {author} {\bibfnamefont {C.}~\bibnamefont {Gorini}},
  \bibinfo {author} {\bibfnamefont {G.}~\bibnamefont {Fleury}}, \ and\ \bibinfo
  {author} {\bibfnamefont {J.-L.}\ \bibnamefont {Pichard}},\ }\href@noop {}
  {\bibfield  {journal} {\bibinfo  {journal} {Phys. Rev. Appl.}\ }\textbf
  {\bibinfo {volume} {3}},\ \bibinfo {pages} {054002} (\bibinfo {year}
  {2015}{\natexlab{b}})}\BibitemShut {NoStop}%
\bibitem [{\citenamefont {Bosisio}\ \emph
  {et~al.}(2015{\natexlab{c}})\citenamefont {Bosisio}, \citenamefont {Gorini},
  \citenamefont {Fleury},\ and\ \citenamefont {Pichard}}]{Bosisio2015bis}%
  \BibitemOpen
  \bibfield  {author} {\bibinfo {author} {\bibfnamefont {R.}~\bibnamefont
  {Bosisio}}, \bibinfo {author} {\bibfnamefont {C.}~\bibnamefont {Gorini}},
  \bibinfo {author} {\bibfnamefont {G.}~\bibnamefont {Fleury}}, \ and\ \bibinfo
  {author} {\bibfnamefont {J.-L.}\ \bibnamefont {Pichard}},\ }\href {\doibase
  http://dx.doi.org/10.1016/j.physe.2015.07.012} {\bibfield  {journal}
  {\bibinfo  {journal} {Physica E: Low-dimensional Systems and Nanostructures}\
  }\textbf {\bibinfo {volume} {74}},\ \bibinfo {pages} {340} (\bibinfo {year}
  {2015}{\natexlab{c}})}\BibitemShut {NoStop}%
\bibitem [{\citenamefont {Pekola}\ and\ \citenamefont
  {Hekking}(2007)}]{Pekola2007}%
  \BibitemOpen
  \bibfield  {author} {\bibinfo {author} {\bibfnamefont {J.~P.}\ \bibnamefont
  {Pekola}}\ and\ \bibinfo {author} {\bibfnamefont {F.~W.~J.}\ \bibnamefont
  {Hekking}},\ }\href@noop {} {\bibfield  {journal} {\bibinfo  {journal} {Phys.
  Rev. Lett.}\ }\textbf {\bibinfo {volume} {98}},\ \bibinfo {pages} {210604}
  (\bibinfo {year} {2007})}\BibitemShut {NoStop}%
\bibitem [{\citenamefont {Jiang}\ \emph
  {et~al.}(2013{\natexlab{b}})\citenamefont {Jiang}, \citenamefont
  {Entin-Wohlman},\ and\ \citenamefont {Imry}}]{Jiang2013}%
  \BibitemOpen
  \bibfield  {author} {\bibinfo {author} {\bibfnamefont {J.-H.}\ \bibnamefont
  {Jiang}}, \bibinfo {author} {\bibfnamefont {O.}~\bibnamefont
  {Entin-Wohlman}}, \ and\ \bibinfo {author} {\bibfnamefont {Y.}~\bibnamefont
  {Imry}},\ }\href@noop {} {\bibfield  {journal} {\bibinfo  {journal} {Phys.
  Rev. B}\ }\textbf {\bibinfo {volume} {87}},\ \bibinfo {pages} {205420}
  (\bibinfo {year} {2013}{\natexlab{b}})}\BibitemShut {NoStop}%
\bibitem [{\citenamefont {Li}\ \emph {et~al.}(2012)\citenamefont {Li},
  \citenamefont {Sun}, \citenamefont {Yao}, \citenamefont {Zhu},\ and\
  \citenamefont {Lu}}]{Li2012}%
  \BibitemOpen
  \bibfield  {author} {\bibinfo {author} {\bibfnamefont {Z.}~\bibnamefont
  {Li}}, \bibinfo {author} {\bibfnamefont {Q.}~\bibnamefont {Sun}}, \bibinfo
  {author} {\bibfnamefont {X.~D.}\ \bibnamefont {Yao}}, \bibinfo {author}
  {\bibfnamefont {Z.~H.}\ \bibnamefont {Zhu}}, \ and\ \bibinfo {author}
  {\bibfnamefont {G.~Q.~M.}\ \bibnamefont {Lu}},\ }\href {\doibase
  10.1039/C2JM33899H} {\bibfield  {journal} {\bibinfo  {journal} {J. Mater.
  Chem.}\ }\textbf {\bibinfo {volume} {22}},\ \bibinfo {pages} {22821}
  (\bibinfo {year} {2012})}\BibitemShut {NoStop}%
\bibitem [{\citenamefont {Kim}\ \emph {et~al.}(2013)\citenamefont {Kim},
  \citenamefont {Bahk}, \citenamefont {Hwang}, \citenamefont {Kim},
  \citenamefont {Park},\ and\ \citenamefont {Kim}}]{Kim2013}%
  \BibitemOpen
  \bibfield  {author} {\bibinfo {author} {\bibfnamefont {J.}~\bibnamefont
  {Kim}}, \bibinfo {author} {\bibfnamefont {J.-H.}\ \bibnamefont {Bahk}},
  \bibinfo {author} {\bibfnamefont {J.}~\bibnamefont {Hwang}}, \bibinfo
  {author} {\bibfnamefont {H.}~\bibnamefont {Kim}}, \bibinfo {author}
  {\bibfnamefont {H.}~\bibnamefont {Park}}, \ and\ \bibinfo {author}
  {\bibfnamefont {W.}~\bibnamefont {Kim}},\ }\href {\doibase
  10.1002/pssr.201307239} {\bibfield  {journal} {\bibinfo  {journal} {Phys.
  Status Solidi RRL}\ }\textbf {\bibinfo {volume} {7}},\ \bibinfo {pages} {767}
  (\bibinfo {year} {2013})}\BibitemShut {NoStop}%
\bibitem [{\citenamefont {Nakpathomkun}\ \emph {et~al.}(2010)\citenamefont
  {Nakpathomkun}, \citenamefont {Xu},\ and\ \citenamefont
  {Linke}}]{Nakpathomkun2010}%
  \BibitemOpen
  \bibfield  {author} {\bibinfo {author} {\bibfnamefont {N.}~\bibnamefont
  {Nakpathomkun}}, \bibinfo {author} {\bibfnamefont {H.~Q.}\ \bibnamefont
  {Xu}}, \ and\ \bibinfo {author} {\bibfnamefont {H.}~\bibnamefont {Linke}},\
  }\href {\doibase 10.1103/PhysRevB.82.235428} {\bibfield  {journal} {\bibinfo
  {journal} {Phys. Rev. B}\ }\textbf {\bibinfo {volume} {82}},\ \bibinfo
  {pages} {235428} (\bibinfo {year} {2010})}\BibitemShut {NoStop}%
\bibitem [{\citenamefont {Hochbaum}\ \emph {et~al.}(2008)\citenamefont
  {Hochbaum}, \citenamefont {Chen}, \citenamefont {Delgado}, \citenamefont
  {Liang}, \citenamefont {Garnett}, \citenamefont {Najarian}, \citenamefont
  {Majumdar},\ and\ \citenamefont {Yang}}]{Hochbaum2008}%
  \BibitemOpen
  \bibfield  {author} {\bibinfo {author} {\bibfnamefont {A.~I.}\ \bibnamefont
  {Hochbaum}}, \bibinfo {author} {\bibfnamefont {R.}~\bibnamefont {Chen}},
  \bibinfo {author} {\bibfnamefont {R.~D.}\ \bibnamefont {Delgado}}, \bibinfo
  {author} {\bibfnamefont {W.}~\bibnamefont {Liang}}, \bibinfo {author}
  {\bibfnamefont {E.~C.}\ \bibnamefont {Garnett}}, \bibinfo {author}
  {\bibfnamefont {M.}~\bibnamefont {Najarian}}, \bibinfo {author}
  {\bibfnamefont {A.}~\bibnamefont {Majumdar}}, \ and\ \bibinfo {author}
  {\bibfnamefont {P.}~\bibnamefont {Yang}},\ }\href@noop {} {\bibfield
  {journal} {\bibinfo  {journal} {Nature}\ }\textbf {\bibinfo {volume} {451}},\
  \bibinfo {pages} {163} (\bibinfo {year} {2008})}\BibitemShut {NoStop}%
\bibitem [{\citenamefont {Persson}\ \emph {et~al.}(2009)\citenamefont
  {Persson}, \citenamefont {Fr\"oberg}, \citenamefont {Samuelson},\ and\
  \citenamefont {Linke}}]{Persson2009}%
  \BibitemOpen
  \bibfield  {author} {\bibinfo {author} {\bibfnamefont {A.~I.}\ \bibnamefont
  {Persson}}, \bibinfo {author} {\bibfnamefont {L.~E.}\ \bibnamefont
  {Fr\"oberg}}, \bibinfo {author} {\bibfnamefont {L.}~\bibnamefont
  {Samuelson}}, \ and\ \bibinfo {author} {\bibfnamefont {H.}~\bibnamefont
  {Linke}},\ }\href@noop {} {\bibfield  {journal} {\bibinfo  {journal}
  {Nanotechnology}\ }\textbf {\bibinfo {volume} {20}},\ \bibinfo {pages}
  {225304} (\bibinfo {year} {2009})}\BibitemShut {NoStop}%
\bibitem [{\citenamefont {Wang}\ and\ \citenamefont {Gates}(2009)}]{Wang2009}%
  \BibitemOpen
  \bibfield  {author} {\bibinfo {author} {\bibfnamefont {M.~C.}\ \bibnamefont
  {Wang}}\ and\ \bibinfo {author} {\bibfnamefont {B.~D.}\ \bibnamefont
  {Gates}},\ }\href@noop {} {\bibfield  {journal} {\bibinfo  {journal}
  {Materials Today}\ }\textbf {\bibinfo {volume} {12}},\ \bibinfo {pages} {34}
  (\bibinfo {year} {2009})}\BibitemShut {NoStop}%
\bibitem [{\citenamefont {Curtin}\ \emph {et~al.}(2012)\citenamefont {Curtin},
  \citenamefont {Fang},\ and\ \citenamefont {Bowers}}]{Curtin2012}%
  \BibitemOpen
  \bibfield  {author} {\bibinfo {author} {\bibfnamefont {B.~M.}\ \bibnamefont
  {Curtin}}, \bibinfo {author} {\bibfnamefont {E.~W.}\ \bibnamefont {Fang}}, \
  and\ \bibinfo {author} {\bibfnamefont {J.~E.}\ \bibnamefont {Bowers}},\
  }\href@noop {} {\bibfield  {journal} {\bibinfo  {journal} {J. Electron.
  Mater.}\ }\textbf {\bibinfo {volume} {41}},\ \bibinfo {pages} {887} (\bibinfo
  {year} {2012})}\BibitemShut {NoStop}%
\bibitem [{\citenamefont {Farrell}\ \emph {et~al.}(2012)\citenamefont
  {Farrell}, \citenamefont {Kinahan}, \citenamefont {Hansel}, \citenamefont
  {Stuen}, \citenamefont {Petkov}, \citenamefont {Shaw}, \citenamefont {West},
  \citenamefont {Djara}, \citenamefont {Dunne}, \citenamefont {Varona},
  \citenamefont {Gleeson}, \citenamefont {S.-J}, \citenamefont {Kim},
  \citenamefont {Kole\'{s}nik}, \citenamefont {Lutz}, \citenamefont {Murray},
  \citenamefont {Holmes}, \citenamefont {Nealey}, \citenamefont {Duesberg},
  \citenamefont {Krsti\'{c}},\ and\ \citenamefont {Morris}}]{Farrell2012}%
  \BibitemOpen
  \bibfield  {author} {\bibinfo {author} {\bibfnamefont {R.~A.}\ \bibnamefont
  {Farrell}}, \bibinfo {author} {\bibfnamefont {N.~T.}\ \bibnamefont
  {Kinahan}}, \bibinfo {author} {\bibfnamefont {S.}~\bibnamefont {Hansel}},
  \bibinfo {author} {\bibfnamefont {K.~O.}\ \bibnamefont {Stuen}}, \bibinfo
  {author} {\bibfnamefont {N.}~\bibnamefont {Petkov}}, \bibinfo {author}
  {\bibfnamefont {M.~T.}\ \bibnamefont {Shaw}}, \bibinfo {author}
  {\bibfnamefont {L.~E.}\ \bibnamefont {West}}, \bibinfo {author}
  {\bibfnamefont {V.}~\bibnamefont {Djara}}, \bibinfo {author} {\bibfnamefont
  {R.~J.}\ \bibnamefont {Dunne}}, \bibinfo {author} {\bibfnamefont {O.~G.}\
  \bibnamefont {Varona}}, \bibinfo {author} {\bibfnamefont {P.~G.}\
  \bibnamefont {Gleeson}}, \bibinfo {author} {\bibfnamefont {J.}~\bibnamefont
  {S.-J}}, \bibinfo {author} {\bibfnamefont {H.-Y.}\ \bibnamefont {Kim}},
  \bibinfo {author} {\bibfnamefont {M.~M.}\ \bibnamefont {Kole\'{s}nik}},
  \bibinfo {author} {\bibfnamefont {T.}~\bibnamefont {Lutz}}, \bibinfo {author}
  {\bibfnamefont {C.~P.}\ \bibnamefont {Murray}}, \bibinfo {author}
  {\bibfnamefont {J.~D.}\ \bibnamefont {Holmes}}, \bibinfo {author}
  {\bibfnamefont {P.~F.}\ \bibnamefont {Nealey}}, \bibinfo {author}
  {\bibfnamefont {G.~S.}\ \bibnamefont {Duesberg}}, \bibinfo {author}
  {\bibfnamefont {V.}~\bibnamefont {Krsti\'{c}}}, \ and\ \bibinfo {author}
  {\bibfnamefont {M.~A.}\ \bibnamefont {Morris}},\ }\href@noop {} {\bibfield
  {journal} {\bibinfo  {journal} {Nanoscale}\ }\textbf {\bibinfo {volume}
  {4}},\ \bibinfo {pages} {3228} (\bibinfo {year} {2012})}\BibitemShut
  {NoStop}%
\bibitem [{\citenamefont {Stranz}\ \emph {et~al.}(2013)\citenamefont {Stranz},
  \citenamefont {Waag},\ and\ \citenamefont {Peiner}}]{Stranz2013}%
  \BibitemOpen
  \bibfield  {author} {\bibinfo {author} {\bibfnamefont {A.}~\bibnamefont
  {Stranz}}, \bibinfo {author} {\bibfnamefont {A.}~\bibnamefont {Waag}}, \ and\
  \bibinfo {author} {\bibfnamefont {E.}~\bibnamefont {Peiner}},\ }\href@noop {}
  {\bibfield  {journal} {\bibinfo  {journal} {J. Electron. Mat.}\ }\textbf
  {\bibinfo {volume} {42}},\ \bibinfo {pages} {2233} (\bibinfo {year}
  {2013})}\BibitemShut {NoStop}%
\bibitem [{\citenamefont {Garnett}\ \emph {et~al.}(2011)\citenamefont
  {Garnett}, \citenamefont {Brongersma}, \citenamefont {Cui},\ and\
  \citenamefont {McGehee}}]{Garnett2011}%
  \BibitemOpen
  \bibfield  {author} {\bibinfo {author} {\bibfnamefont {E.~C.}\ \bibnamefont
  {Garnett}}, \bibinfo {author} {\bibfnamefont {M.~L.}\ \bibnamefont
  {Brongersma}}, \bibinfo {author} {\bibfnamefont {Y.}~\bibnamefont {Cui}}, \
  and\ \bibinfo {author} {\bibfnamefont {M.~D.}\ \bibnamefont {McGehee}},\
  }\href {\doibase 10.1146/annurev-matsci-062910-100434} {\bibfield  {journal}
  {\bibinfo  {journal} {Annual Review of Materials Research}\ }\textbf
  {\bibinfo {volume} {41}},\ \bibinfo {pages} {269} (\bibinfo {year}
  {2011})}\BibitemShut {NoStop}%
\bibitem [{\citenamefont {LaPierre}\ \emph {et~al.}(2013)\citenamefont
  {LaPierre}, \citenamefont {Chia}, \citenamefont {Gibson}, \citenamefont
  {Haapamaki}, \citenamefont {Boulanger}, \citenamefont {Yee}, \citenamefont
  {Kuyanov}, \citenamefont {Zhang}, \citenamefont {Tajik}, \citenamefont
  {Jewell},\ and\ \citenamefont {Rahman}}]{Lapierre2013}%
  \BibitemOpen
  \bibfield  {author} {\bibinfo {author} {\bibfnamefont {R.~R.}\ \bibnamefont
  {LaPierre}}, \bibinfo {author} {\bibfnamefont {A.~C.~E.}\ \bibnamefont
  {Chia}}, \bibinfo {author} {\bibfnamefont {S.~J.}\ \bibnamefont {Gibson}},
  \bibinfo {author} {\bibfnamefont {C.~M.}\ \bibnamefont {Haapamaki}}, \bibinfo
  {author} {\bibfnamefont {J.}~\bibnamefont {Boulanger}}, \bibinfo {author}
  {\bibfnamefont {R.}~\bibnamefont {Yee}}, \bibinfo {author} {\bibfnamefont
  {P.}~\bibnamefont {Kuyanov}}, \bibinfo {author} {\bibfnamefont
  {J.}~\bibnamefont {Zhang}}, \bibinfo {author} {\bibfnamefont
  {N.}~\bibnamefont {Tajik}}, \bibinfo {author} {\bibfnamefont
  {N.}~\bibnamefont {Jewell}}, \ and\ \bibinfo {author} {\bibfnamefont
  {K.~M.~A.}\ \bibnamefont {Rahman}},\ }\href {\doibase 10.1002/pssr.201307109}
  {\bibfield  {journal} {\bibinfo  {journal} {physica status solidi (RRL) -
  Rapid Research Letters}\ }\textbf {\bibinfo {volume} {7}},\ \bibinfo {pages}
  {815} (\bibinfo {year} {2013})}\BibitemShut {NoStop}%
\bibitem [{\citenamefont {Chen}\ \emph {et~al.}(2011)\citenamefont {Chen},
  \citenamefont {Li},\ and\ \citenamefont {Chen}}]{Chen2011}%
  \BibitemOpen
  \bibfield  {author} {\bibinfo {author} {\bibfnamefont {K.-I.}\ \bibnamefont
  {Chen}}, \bibinfo {author} {\bibfnamefont {B.-R.}\ \bibnamefont {Li}}, \ and\
  \bibinfo {author} {\bibfnamefont {Y.-T.}\ \bibnamefont {Chen}},\ }\href
  {\doibase http://dx.doi.org/10.1016/j.nantod.2011.02.001} {\bibfield
  {journal} {\bibinfo  {journal} {Nano Today}\ }\textbf {\bibinfo {volume}
  {6}},\ \bibinfo {pages} {131} (\bibinfo {year} {2011})}\BibitemShut {NoStop}%
\bibitem [{\citenamefont {Mott}(1969)}]{Mott1969}%
  \BibitemOpen
  \bibfield  {author} {\bibinfo {author} {\bibfnamefont {N.~F.}\ \bibnamefont
  {Mott}},\ }\href@noop {} {\bibfield  {journal} {\bibinfo  {journal} {Phil.
  Mag.}\ }\textbf {\bibinfo {volume} {19}},\ \bibinfo {pages} {835} (\bibinfo
  {year} {1969})}\BibitemShut {NoStop}%
\bibitem [{\citenamefont {Ambegaokar}\ \emph {et~al.}(1971)\citenamefont
  {Ambegaokar}, \citenamefont {Halperin},\ and\ \citenamefont
  {Langer}}]{Ambegaokar1971}%
  \BibitemOpen
  \bibfield  {author} {\bibinfo {author} {\bibfnamefont {V.}~\bibnamefont
  {Ambegaokar}}, \bibinfo {author} {\bibfnamefont {B.~I.}\ \bibnamefont
  {Halperin}}, \ and\ \bibinfo {author} {\bibfnamefont {J.~S.}\ \bibnamefont
  {Langer}},\ }\href@noop {} {\bibfield  {journal} {\bibinfo  {journal} {Phys.
  Rev. B}\ }\textbf {\bibinfo {volume} {4}},\ \bibinfo {pages} {2612} (\bibinfo
  {year} {1971})}\BibitemShut {NoStop}%
\bibitem [{\citenamefont {Rahman}\ \emph {et~al.}(2006)\citenamefont {Rahman},
  \citenamefont {Sanyal}, \citenamefont {Gangopadhayy}, \citenamefont {De},\
  and\ \citenamefont {Das}}]{Rahman2006}%
  \BibitemOpen
  \bibfield  {author} {\bibinfo {author} {\bibfnamefont {A.}~\bibnamefont
  {Rahman}}, \bibinfo {author} {\bibfnamefont {M.~K.}\ \bibnamefont {Sanyal}},
  \bibinfo {author} {\bibfnamefont {R.}~\bibnamefont {Gangopadhayy}}, \bibinfo
  {author} {\bibfnamefont {A.}~\bibnamefont {De}}, \ and\ \bibinfo {author}
  {\bibfnamefont {I.}~\bibnamefont {Das}},\ }\href {\doibase
  10.1103/PhysRevB.73.125313} {\bibfield  {journal} {\bibinfo  {journal} {Phys.
  Rev. B}\ }\textbf {\bibinfo {volume} {73}},\ \bibinfo {pages} {125313}
  (\bibinfo {year} {2006})}\BibitemShut {NoStop}%
\bibitem [{\citenamefont {Linke}\ \emph {et~al.}(1999)\citenamefont {Linke},
  \citenamefont {Humphrey}, \citenamefont {L\"ofgren}, \citenamefont {Sushkov},
  \citenamefont {Newbury}, \citenamefont {Taylor},\ and\ \citenamefont
  {Omling}}]{Linke1999}%
  \BibitemOpen
  \bibfield  {author} {\bibinfo {author} {\bibfnamefont {H.}~\bibnamefont
  {Linke}}, \bibinfo {author} {\bibfnamefont {T.~E.}\ \bibnamefont {Humphrey}},
  \bibinfo {author} {\bibfnamefont {A.}~\bibnamefont {L\"ofgren}}, \bibinfo
  {author} {\bibfnamefont {A.~O.}\ \bibnamefont {Sushkov}}, \bibinfo {author}
  {\bibfnamefont {R.}~\bibnamefont {Newbury}}, \bibinfo {author} {\bibfnamefont
  {R.~P.}\ \bibnamefont {Taylor}}, \ and\ \bibinfo {author} {\bibfnamefont
  {P.}~\bibnamefont {Omling}},\ }\href {\doibase 10.1126/science.286.5448.2314}
  {\bibfield  {journal} {\bibinfo  {journal} {Science}\ }\textbf {\bibinfo
  {volume} {286}},\ \bibinfo {pages} {2314} (\bibinfo {year}
  {1999})}\BibitemShut {NoStop}%
\bibitem [{\citenamefont {Sassine}\ \emph {et~al.}(2008)\citenamefont
  {Sassine}, \citenamefont {Krupko}, \citenamefont {Portal}, \citenamefont
  {Kvon}, \citenamefont {Murali}, \citenamefont {Martin}, \citenamefont
  {Hill},\ and\ \citenamefont {Wieck}}]{Sassine2008}%
  \BibitemOpen
  \bibfield  {author} {\bibinfo {author} {\bibfnamefont {S.}~\bibnamefont
  {Sassine}}, \bibinfo {author} {\bibfnamefont {Y.}~\bibnamefont {Krupko}},
  \bibinfo {author} {\bibfnamefont {J.-C.}\ \bibnamefont {Portal}}, \bibinfo
  {author} {\bibfnamefont {Z.~D.}\ \bibnamefont {Kvon}}, \bibinfo {author}
  {\bibfnamefont {R.}~\bibnamefont {Murali}}, \bibinfo {author} {\bibfnamefont
  {K.~P.}\ \bibnamefont {Martin}}, \bibinfo {author} {\bibfnamefont
  {G.}~\bibnamefont {Hill}}, \ and\ \bibinfo {author} {\bibfnamefont {A.~D.}\
  \bibnamefont {Wieck}},\ }\href {\doibase 10.1103/PhysRevB.78.045431}
  {\bibfield  {journal} {\bibinfo  {journal} {Phys. Rev. B}\ }\textbf {\bibinfo
  {volume} {78}},\ \bibinfo {pages} {045431} (\bibinfo {year}
  {2008})}\BibitemShut {NoStop}%
\bibitem [{\citenamefont {Callen}(1985)}]{Callen1985}%
  \BibitemOpen
  \bibfield  {author} {\bibinfo {author} {\bibfnamefont {H.}~\bibnamefont
  {Callen}},\ }\href@noop {} {\emph {\bibinfo {title} {Thermodynamics and an
  Introduction to Thermostatics}}}\ (\bibinfo  {publisher} {John Wiley and
  Sons, New York},\ \bibinfo {year} {1985})\BibitemShut {NoStop}%
\bibitem [{Note1()}]{Note1}%
  \BibitemOpen
  \bibinfo {note} {In previous works~\cite
  {Bosisio20142,BosisioGorini2015,Bosisio2015bis} focusing on band-edge
  transport, the energy dependence of the localization length was crucial and
  thus taken into account. Such dependence is here largely inconsenquential:
  apart from a brief discussion of the less relevant configuration of Fig.~\ref
  {fig:ratchet}(a1), (a2), (a3), the band edges will not be
  probed.}\BibitemShut {Stop}%
\bibitem [{\citenamefont {Miller}\ and\ \citenamefont
  {Abrahams}(1960)}]{Miller1960}%
  \BibitemOpen
  \bibfield  {author} {\bibinfo {author} {\bibfnamefont {A.}~\bibnamefont
  {Miller}}\ and\ \bibinfo {author} {\bibfnamefont {E.}~\bibnamefont
  {Abrahams}},\ }\href@noop {} {\bibfield  {journal} {\bibinfo  {journal}
  {Phys. Rev.}\ }\textbf {\bibinfo {volume} {120}},\ \bibinfo {pages} {745}
  (\bibinfo {year} {1960})}\BibitemShut {NoStop}%
\bibitem [{Note2()}]{Note2}%
  \BibitemOpen
  \bibinfo {note} {It is indeed close to the Mott temperature $k_BT_M=2/\xi \nu
  =2\protect \tmspace +\thinmuskip {.1667em}\epsilon $ and much larger than the
  activation temperature $k_BT_x=\xi /(2\nu N^2)=8.10^{-4}\protect \tmspace
  +\thinmuskip {.1667em}\epsilon $ (see Ref.\protect \rev@citealpnum
  {Bosisio20142} for more details).}\BibitemShut {Stop}%
\bibitem [{\citenamefont {Feynman}\ \emph {et~al.}(1964)\citenamefont
  {Feynman}, \citenamefont {Leighton},\ and\ \citenamefont
  {Sands}}]{Feynman1964}%
  \BibitemOpen
  \bibfield  {author} {\bibinfo {author} {\bibfnamefont {R.~P.}\ \bibnamefont
  {Feynman}}, \bibinfo {author} {\bibfnamefont {B.~R.}\ \bibnamefont
  {Leighton}}, \ and\ \bibinfo {author} {\bibfnamefont {M.}~\bibnamefont
  {Sands}},\ }\href@noop {} {\emph {\bibinfo {title} {The Feynman Lectures on
  Physics}}},\ Vol.\ \bibinfo {volume} {1.46}\ (\bibinfo  {publisher} {Addison
  - Wesley},\ \bibinfo {year} {1964})\BibitemShut {NoStop}%
\bibitem [{Note3()}]{Note3}%
  \BibitemOpen
  \bibinfo {note} {It is worth to stress that if we consider a single NW,
  electron-hole symmetry may be broken even at $\mu =0$ due to disorder;
  however, when considering a large set of NWs having constant density of
  states $\nu $, symmetry is restored on average.}\BibitemShut {Stop}%
\bibitem [{\citenamefont {Bosisio}\ \emph
  {et~al.}(2014{\natexlab{b}})\citenamefont {Bosisio}, \citenamefont {Fleury},\
  and\ \citenamefont {Pichard}}]{Bosisio20141}%
  \BibitemOpen
  \bibfield  {author} {\bibinfo {author} {\bibfnamefont {R.}~\bibnamefont
  {Bosisio}}, \bibinfo {author} {\bibfnamefont {G.}~\bibnamefont {Fleury}}, \
  and\ \bibinfo {author} {\bibfnamefont {J.-L.}\ \bibnamefont {Pichard}},\
  }\href@noop {} {\bibfield  {journal} {\bibinfo  {journal} {New J. Phys.}\
  }\textbf {\bibinfo {volume} {16}},\ \bibinfo {pages} {035004} (\bibinfo
  {year} {2014}{\natexlab{b}})}\BibitemShut {NoStop}%
\bibitem [{Note4()}]{Note4}%
  \BibitemOpen
  \bibinfo {note} {They give an estimation of the difference between data
  obtained for finite $M$ and the quantity $I^N_L/M$ for $M\to \infty
  $.}\BibitemShut {Stop}%
\bibitem [{Note5()}]{Note5}%
  \BibitemOpen
  \bibinfo {note} {We refer to the ``electronic'' figure of merit $ZT$ to
  distinguish it from the ``full'' figure of merit $\protect \overline {ZT}$,
  which would include also the phononic contributions to the thermal
  conductance, here neglected.}\BibitemShut {Stop}%
\bibitem [{Note6()}]{Note6}%
  \BibitemOpen
  \bibinfo {note} {These coefficients are all special instances of the more
  general ones discussed in Ref.~\protect \rev@citealpnum
  {Mazza2014}.}\BibitemShut {Stop}%
\bibitem [{\citenamefont {Benenti}\ \emph {et~al.}(2013)\citenamefont
  {Benenti}, \citenamefont {Casati}, \citenamefont {Prosen},\ and\
  \citenamefont {Saito}}]{Benenti2013}%
  \BibitemOpen
  \bibfield  {author} {\bibinfo {author} {\bibfnamefont {G.}~\bibnamefont
  {Benenti}}, \bibinfo {author} {\bibfnamefont {G.}~\bibnamefont {Casati}},
  \bibinfo {author} {\bibfnamefont {T.}~\bibnamefont {Prosen}}, \ and\ \bibinfo
  {author} {\bibfnamefont {K.}~\bibnamefont {Saito}},\ }\href@noop {}
  {\bibfield  {journal} {\bibinfo  {journal} {arXiv:1311.4430}\ } (\bibinfo
  {year} {2013})}\BibitemShut {NoStop}%
\bibitem [{Note7()}]{Note7}%
  \BibitemOpen
  \bibinfo {note} {Switching from the single barrier configuration to the
  double barrier configuration, the Onsager coefficient $L_{13}$ is increased
  while $L_{11}$ is decreased, in such a way that the ratio $L_{13}/L_{11}$ is
  exactly doubled for any value of $E_s$.}\BibitemShut {Stop}%
\bibitem [{\citenamefont {Curzon.}\ and\ \citenamefont
  {Ahlborn}(1975)}]{CurzonAhlborn1975}%
  \BibitemOpen
  \bibfield  {author} {\bibinfo {author} {\bibfnamefont {F.}~\bibnamefont
  {Curzon.}}\ and\ \bibinfo {author} {\bibfnamefont {B.}~\bibnamefont
  {Ahlborn}},\ }\href@noop {} {\bibfield  {journal} {\bibinfo  {journal} {Am.
  J. Phys.}\ }\textbf {\bibinfo {volume} {43}},\ \bibinfo {pages} {22}
  (\bibinfo {year} {1975})}\BibitemShut {NoStop}%
\bibitem [{Note8()}]{Note8}%
  \BibitemOpen
  \bibinfo {note} {It is also interesting to give an estimation of the
  reachable output power density. It depends on the geometric parameters of the
  NW array. They must be chosen so as to be consistent with the experimental
  constraints and with the 1D model used in this work (typically, the NW
  diameter needs to be smaller than the Mott hopping length\cite {Bosisio20142}
  $L_M\approx \protect \sqrt {\xi /(2\nu k_BT)}\approx 4a$). Also, since the
  reported results are mostly independent of the NW length $L$ as long as $L\gg
  L_M$ is long enough to be in the Mott hopping regime, choosing a short $L$ is
  preferable. By considering for instance a NW diameter of $20\protect \tmspace
  +\thinmuskip {.1667em}\protect \mathrm {nm}$ and a packing density of $20\%$,
  $M=10^5$ NWs can be stacked in parallel along a $1\protect \tmspace
  +\thinmuskip {.1667em}\protect \mathrm {cm}$ wide chip, yielding $P_{\protect
  \text {max}}\approx 1\protect \tmspace +\thinmuskip {.1667em}\mu \protect
  \mathrm {W}$ (keeping the same model parameters as right above). Taking
  $1\protect \tmspace +\thinmuskip {.1667em}\mu \protect \mathrm {m}$-long NWs,
  this gives an output power density of the order of $10\protect \tmspace
  +\thinmuskip {.1667em}\protect \mathrm {mW/cm}^2$.}\BibitemShut {Stop}%
\bibitem [{Note9()}]{Note9}%
  \BibitemOpen
  \bibinfo {note} {Additional heat exchanges between the substrate phonons and
  the electrodes could take place via direct contact between them, or via
  phonon-mediated processes involving NWs phonons.}\BibitemShut {Stop}%
\bibitem [{\citenamefont {Venkatasubramanian}(2000)}]{Venkatasubramanian2000}%
  \BibitemOpen
  \bibfield  {author} {\bibinfo {author} {\bibfnamefont {R.}~\bibnamefont
  {Venkatasubramanian}},\ }\href@noop {} {\bibfield  {journal} {\bibinfo
  {journal} {Phys. Rev. B}\ }\textbf {\bibinfo {volume} {61}},\ \bibinfo
  {pages} {3091} (\bibinfo {year} {2000})}\BibitemShut {NoStop}%
\bibitem [{\citenamefont {Heron}\ \emph {et~al.}(2010)\citenamefont {Heron},
  \citenamefont {Bera}, \citenamefont {Fournier}, \citenamefont {Mingo},\ and\
  \citenamefont {Bourgeois}}]{Heron2010}%
  \BibitemOpen
  \bibfield  {author} {\bibinfo {author} {\bibfnamefont {J.-S.}\ \bibnamefont
  {Heron}}, \bibinfo {author} {\bibfnamefont {C.}~\bibnamefont {Bera}},
  \bibinfo {author} {\bibfnamefont {T.}~\bibnamefont {Fournier}}, \bibinfo
  {author} {\bibfnamefont {N.}~\bibnamefont {Mingo}}, \ and\ \bibinfo {author}
  {\bibfnamefont {O.}~\bibnamefont {Bourgeois}},\ }\href@noop {} {\bibfield
  {journal} {\bibinfo  {journal} {Phys. Rev. B}\ }\textbf {\bibinfo {volume}
  {82}},\ \bibinfo {pages} {155458} (\bibinfo {year} {2010})}\BibitemShut
  {NoStop}%
\bibitem [{\citenamefont {Yu}\ \emph {et~al.}(2010)\citenamefont {Yu},
  \citenamefont {Mitrovic}, \citenamefont {Tham}, \citenamefont {Varghese},\
  and\ \citenamefont {Heath}}]{Yu2010}%
  \BibitemOpen
  \bibfield  {author} {\bibinfo {author} {\bibfnamefont {J.-K.}\ \bibnamefont
  {Yu}}, \bibinfo {author} {\bibfnamefont {S.}~\bibnamefont {Mitrovic}},
  \bibinfo {author} {\bibfnamefont {D.}~\bibnamefont {Tham}}, \bibinfo {author}
  {\bibfnamefont {J.}~\bibnamefont {Varghese}}, \ and\ \bibinfo {author}
  {\bibfnamefont {J.~R.}\ \bibnamefont {Heath}},\ }\href@noop {} {\bibfield
  {journal} {\bibinfo  {journal} {Nat. Nanotechnology}\ }\textbf {\bibinfo
  {volume} {5}},\ \bibinfo {pages} {718} (\bibinfo {year} {2010})}\BibitemShut
  {NoStop}%
\bibitem [{\citenamefont {He}\ and\ \citenamefont {Galli}(2012)}]{He2012}%
  \BibitemOpen
  \bibfield  {author} {\bibinfo {author} {\bibfnamefont {Y.}~\bibnamefont
  {He}}\ and\ \bibinfo {author} {\bibfnamefont {G.}~\bibnamefont {Galli}},\
  }\href@noop {} {\bibfield  {journal} {\bibinfo  {journal} {Phys. Rev. Lett.}\
  }\textbf {\bibinfo {volume} {108}},\ \bibinfo {pages} {215901} (\bibinfo
  {year} {2012})}\BibitemShut {NoStop}%
\bibitem [{\citenamefont {Denisov}\ \emph {et~al.}(2014)\citenamefont
  {Denisov}, \citenamefont {Flach},\ and\ \citenamefont
  {H\"{a}nggi}}]{Denisov2014}%
  \BibitemOpen
  \bibfield  {author} {\bibinfo {author} {\bibfnamefont {S.}~\bibnamefont
  {Denisov}}, \bibinfo {author} {\bibfnamefont {S.}~\bibnamefont {Flach}}, \
  and\ \bibinfo {author} {\bibfnamefont {P.}~\bibnamefont {H\"{a}nggi}},\
  }\href {\doibase doi:10.1016/j.physrep.2014.01.003} {\bibfield  {journal}
  {\bibinfo  {journal} {Phys. Rep.}\ }\textbf {\bibinfo {volume} {538}},\
  \bibinfo {pages} {77} (\bibinfo {year} {2014})}\BibitemShut {NoStop}%
\end{thebibliography}%
\end{document}